\documentclass[aps,10pt,prb,superscriptaddress,longbibliography,twocolumn]{revtex4-2}
\usepackage[utf8]{inputenc}
\usepackage[T1]{fontenc}

\usepackage{graphicx} 
\usepackage{bm} 
\usepackage{color}
\usepackage{comment}
\usepackage{verbatim}
\usepackage{soul}
\usepackage{placeins}

\begin{document}
\date{\today}
\def\nacosbo{$\mathrm{Na_3Co_2SbO_6}$~}
\title{Pressure tuning of Kitaev spin liquid candidate Na$_3$Co$_2$SbO$_6$}

\author{E. H. T. Poldi}
    \affiliation{Department of Physics, University of Illinois at Chicago, Chicago, Illinois 60607, USA}
    \affiliation{Advanced Photon Source, Argonne National Laboratory, Argonne, Illinois 60439, USA}
\author{R. Tartaglia}
    \affiliation{Institute of Physics Gleb Wataghin, University of Campinas, Campinas, Sao Paulo 13083-859, Brazil}
    \affiliation{Advanced Photon Source, Argonne National Laboratory, Argonne, Illinois 60439, USA}
\author{G. Fabbris}
    \affiliation{Advanced Photon Source, Argonne National Laboratory, Argonne, Illinois 60439, USA}
\author{N. Nguyen}
    \affiliation{Department of Physics, University of Illinois at Chicago, Chicago, Illinois 60607, USA}
\author{H. Park}
    \affiliation{Department of Physics, University of Illinois at Chicago, Chicago, Illinois 60607, USA}
    \affiliation{Materials Science Division, Argonne National Laboratory, Argonne, Illinois 60439, USA}
\author{Z. Liu}
    \affiliation{Department of Physics, University of Illinois at Chicago, Chicago, Illinois 60607, USA}
\author{M. van Veenendaal}
    \affiliation{Department of Physics, Northern Illinois University, DeKalb, Illinois 60115, USA}
\author{R. Kumar}
    \affiliation{Department of Physics, University of Illinois at Chicago, Chicago, Illinois 60607, USA}
\author{G. Jose}
    \affiliation{Department of Physics, University of Alabama at Birmingham, Birmingham, Alabama 35294, USA}
\author{S. Samanta}
    \affiliation{Department of Physics, University of Alabama at Birmingham, Birmingham, Alabama 35294, USA}
\author{W. Bi}
    \affiliation{Department of Physics, University of Alabama at Birmingham, Birmingham, Alabama 35294, USA}
\author{Y. Xiao}
    \affiliation{HPCAT, Advanced Photon Source, Argonne National Laboratory, Argonne, Illinois 60439, USA}
\author{D. Popov}
    \affiliation{HPCAT, Advanced Photon Source, Argonne National Laboratory, Argonne, Illinois 60439, USA}
\author{Y. Wu}
    \affiliation{High Flux Isotope Reactor, Oak Ridge National Laboratory, Oak Ridge, Tennessee 37831, USA}
\author{J.-W. Kim}
    \affiliation{Advanced Photon Source, Argonne National Laboratory, Argonne, Illinois 60439, USA}
\author{H. Zheng}
    \affiliation{Materials Science Division, Argonne National Laboratory, Argonne, Illinois 60439, USA}
\author{J. Yan}
    \affiliation{Materials Science and Technology Division, Oak Ridge National Laboratory, Oak Ridge, Tennessee 37831, USA}
\author{J. F. Mitchell}
    \affiliation{Materials Science Division, Argonne National Laboratory, Argonne, Illinois 60439, USA}
\author{R. J. Hemley}
    \affiliation{Departments of Physics, Chemistry and Earth and Environmental Sciences, University of Illinois at Chicago, Chicago, Illinois 60607, USA}
\author{D. Haskel}
    \email{haskel@anl.gov}
    \affiliation{Advanced Photon Source, Argonne National Laboratory, Argonne, Illinois 60439, USA}


\begin{abstract}

    The search for Kitaev’s quantum spin liquid (KQSL) state in real materials has recently expanded with the prediction that honeycomb lattices of divalent, high-spin cobalt ions could host the dominant bond-dependent exchange interactions required to stabilize the elusive entangled quantum state. The layered honeycomb Na$_3$Co$_2$SbO$_6$ has been singled out as a leading candidate provided that the trigonal crystal field acting on Co $3d$ orbitals, which enhances non-Kitaev exchange interactions between $J_{\rm eff}=\frac{1}{2}$ spin-orbital pseudospins, is reduced. We find that applied pressure leads to anisotropic compression of the layered structure, significantly reducing the trigonal distortion of CoO$_6$ octahedra. A strong enhancement of ferromagnetic correlations between pseudospins is observed in the spin-polarized (3 Tesla) phase up to about 60 GPa. Higher pressures drive a spin transition into a low-spin state destroying the $J_{\rm eff}=\frac{1}{2}$ local moments required to map the spin Hamiltonian into Kitaev's model. The spin transition strongly suppresses the low-temperature magnetic susceptibility and appears to stabilize a paramagnetic phase driven by frustration. Although applied pressure fails to realize a KQSL state, the possible emergence of frustrated magnetism of localized, low-spin $S=\frac{1}{2}$ moments opens the door for exploration of novel magnetic quantum states in compressed honeycomb lattices of divalent cobaltates.

\end{abstract}

\maketitle

\thispagestyle{empty}

The quest to realize Kitaev’s theoretical quantum spin liquid (KQSL) state in real materials is partly driven by the prediction that a magnetic field would drive the gapless solution of Kitaev's model into a gapped phase with anyonic excitations amenable for topologically protected quantum computing \cite{Kitaev2006,Kitaev2003,Freedman2002,Nayak2008}. The materials search has up to recently been largely focused on honeycomb lattices of Ru$^{3+}$ and Ir$^{4+}$ transition metal (TM) ions with a $d^5$ configuration, where a large octahedral crystal field rooted in extended $4d/5d$ orbitals stabilizes a low spin $t^5_{2g}e_g^0$ state and where strong spin-orbit coupling entangles orbital $l_{\rm eff}=1$ and spin $s=\frac{1}{2}$ moments into $j_{\rm eff}=\frac{1}{2}$ pseudospins \cite{Kim2008PRL}. The spin-orbital entanglement introduces bond-directional exchange anisotropy which, in perfect honeycomb lattices of edge-shared octahedra, maps the 90-degree TM-O-TM superexchange interactions into Kitaev’s model \cite{Jackeli2009PRL}. Although sizable Kitaev exchange ($K$) interactions are found in a number of two-dimensional $4d$ and $5d$ honeycomb lattices, including the heavily studied RuCl$_3$, A$_2$IrO$_3$ (A = Na, Li), and A$_3$LiIr$_2$O$_6$ (A = H, Ag) \cite{Takagi2019NatureReviews}, with the exception of H$_3$LiIr$_2$O$_6$ all order magnetically as a result of competing isotropic Heisenberg ($J$) and off-diagonal anisotropic ($\Gamma, \Gamma^{\prime})$ exchange interactions. These non-Kitaev superexchange couplings emerge from  distortions away from the ideal honeycomb lattice, such as those driven by trigonal distortions of TMO$_6$ octahedra, which lift the electronic degeneracy of $t_{2g}$ states and modify the spin-orbital pseudospin wavefunction. Other non-Kitaev interactions include direct isotropic exchange from $d-d$ overlap between extended $4d$/$5d$-orbitals.
    
The more localized nature of $3d$ orbitals makes them attractive for suppression of direct exchange terms. A smaller octahedral crystal field and larger intra-atomic Hund's coupling relative to the $4d/5d$ case stabilizes the high-spin (HS), $t^5_{2g}e^2_g$ configuration in Co$^{2+}$ ions. Spin-orbit coupling combines the $l_{\rm eff}=1$ state formed by the single hole in the 3-fold degenerate $t_{2g}$ states with the total $S=\frac{3}{2}$ spin of $t_{2g}$ and $e_g$ orbitals leading to a many body $J_{\rm eff}=\frac{1}{2}$ pseudospin ground state with excited $J_{\rm eff}=\frac{3}{2}, \frac{5}{2}$  multiplets (we use lower case $j$ for single particle states and upper case $J$ for many body states). However, the order-of-magnitude weaker spin-orbit coupling (tens vs hundreds of meV) makes for a delicate spin-orbital pseudospin moment which is fragile against non-cubic crystal fields arising from lattice distortions. Perfect honeycomb lattices of HS Co$^{2+}$ ions have been predicted to host dominant ferromagnetic Kitaev interactions between $J_{\rm eff}=\frac{1}{2}$ pseudospins, primarily arising from $t_{2g}$-$e_g$ superexchange pathways \cite{Liu2020PRL,Liu2018PRB,Sano2018PRB}. Inelastic neutron scattering \cite{Songvilay2020PRB,Yao2022PRL,Kim2022JPCM} and Co $L_{2,3}$-edge x-ray spectroscopy \cite{Veenendaal2023} provided support for a $J_{\rm eff}=\frac{1}{2}$ description of the ground state in the trigonally distorted honeycomb lattices Na$_2$Co$_2$TeO$_6$ and Na$_3$Co$_2$SbO$_6$, based on measurements of spin-orbit excitations and orbital-to-spin moment ratios, respectively. Whether Kitaev interactions are dominant in these and the related BaCo$_2$(AsO$_4$)$_2$ remains a matter of debate, with both $KJ\Gamma\Gamma^{\prime}$ \cite{Songvilay2020PRB,Kim2022JPCM} and XXZ easy-plane \cite{Gu2024PRB,Halloran-PNAS} spin Hamiltonians being proposed to explain the zig-zag-like magnetic order and spectrum of low-energy magnetic excitations. However, in analogy to RuCl$_3$ where a $\sim 7$ T applied magnetic field suppresses magnetic order at a quantum critical point giving rise to a gapped continuum of magnetic excitations resembling those of a  KQSL \cite{Banerjee2016NatureMat,Wolter-PRB}, magnetic fields below 2 T suppress the antiferromagnetic, zig-zag-like double-Q \cite{Gu2024PRB,Viciu2007,Yan2019PRM} magnetic structure of Na$_3$Co$_2$SbO$_6$ leading to a quantum critical regime featuring a magnetically disordered, spin-liquid-like phase before a spin-polarized phase with gapped magnetic excitations emerges at higher fields \cite{Vavilova2023PRB,Hu2024PRB,Li2022PRX}. The much reduced critical fields needed to enter a quantum critical regime in Na$_3$Co$_2$SbO$_6$ and BaCo$_2$(AsO$_4$)$_2$ \cite{Zhang2023NatMat} relative to RuCl$_3$ suggests  weaker non-Kitaev interactions.

Here we explore whether pressure can stabilize a KQSL state. We focus on Na$_3$Co$_2$SbO$_6$, which is predicted to host a KQSL state if the trigonal crystal field can be at least halved \cite{Liu2020PRL,Liu2018PRB}. Applied pressure allows for continuous tuning of interatomic distances, bond angles, and degree of covalency in TM-O bonds, which dictate crystal fields and exchange interactions. The localized nature of $3d$ orbitals renders cobaltates much more robust against dimerization and formation of molecular orbitals between closely spaced TM ions across edge-shared octahedra. Such pressure-induced dimerization is prevalent in $4d/5d$ honeycomb lattices \cite{Takayama2022PRR,Hermann2019PRB,Xu-NIO,Shen2021PRB,Veiga2019PRB,Fabbris2021PRB} and leads to the collapse of their spin-orbital $j_{\rm eff}=\frac{1}{2}$ moments required to stabilize the KQSL state \cite{Veenendaal2022}. On the other hand, the weak spin-orbit coupling and proclivity of HS Co$^{2+}$ ions to undergo spin transitions under increasing octahedral crystal field may render the many body $J_{\rm eff}=\frac{1}{2}$ moment unstable against lattice compression. We find that quasi-hydrostatic pressure leads to anisotropic lattice compression, which, based on density functional theory calculations of relaxed atomic positions, results in a continuous, sizable reduction of the trigonal distortion in CoO$_6$ octahedra. A significant enhancement of ferromagnetic correlations between pseudospins is observed in the spin-polarized phase ($H=3$ T) below about 60 GPa. While Kitaev exchange interactions in HS $3d^7$ cobaltates are always ferromagnetic \cite{Sano2018PRB,Liu2018PRB,Liu2020PRL}, magnetization near full moment values in 3 T applied field that persists to rather high temperatures points to enhanced Heisenberg ferromagnetic interactions between pseudospins. Pressures above about 70 GPa lead to a spin transition to a low-spin (LS) Co $3d^7$ state quenching the orbital degrees of freedom at the root of bond-directional Kitaev's exchange. While pressure does not stabilize a KQSL state, the honeycomb arrangement of LS Co$^{2+}$ ions shows a vanishing low-temperature magnetic susceptibility at 100 GPa suggestive of a paramagnetic ground state driven by frustration, as predicted by $J_1-J_2-J_3$ quantum models of honeycomb lattices \cite{Jiang2023PRB,Fouet-J1J2}. Since the LS Co $3d^7$ configuration is rarely found at ambient conditions, the results highlight the potential of high-pressure studies to unravel novel frustrated magnetic states in cobaltate honeycomb lattices.

\section*{Results}

\subsection*{Crystal structure and trigonal distortion}
    
\begin{figure*}[ht]
    \centering
    \includegraphics[width=\textwidth]{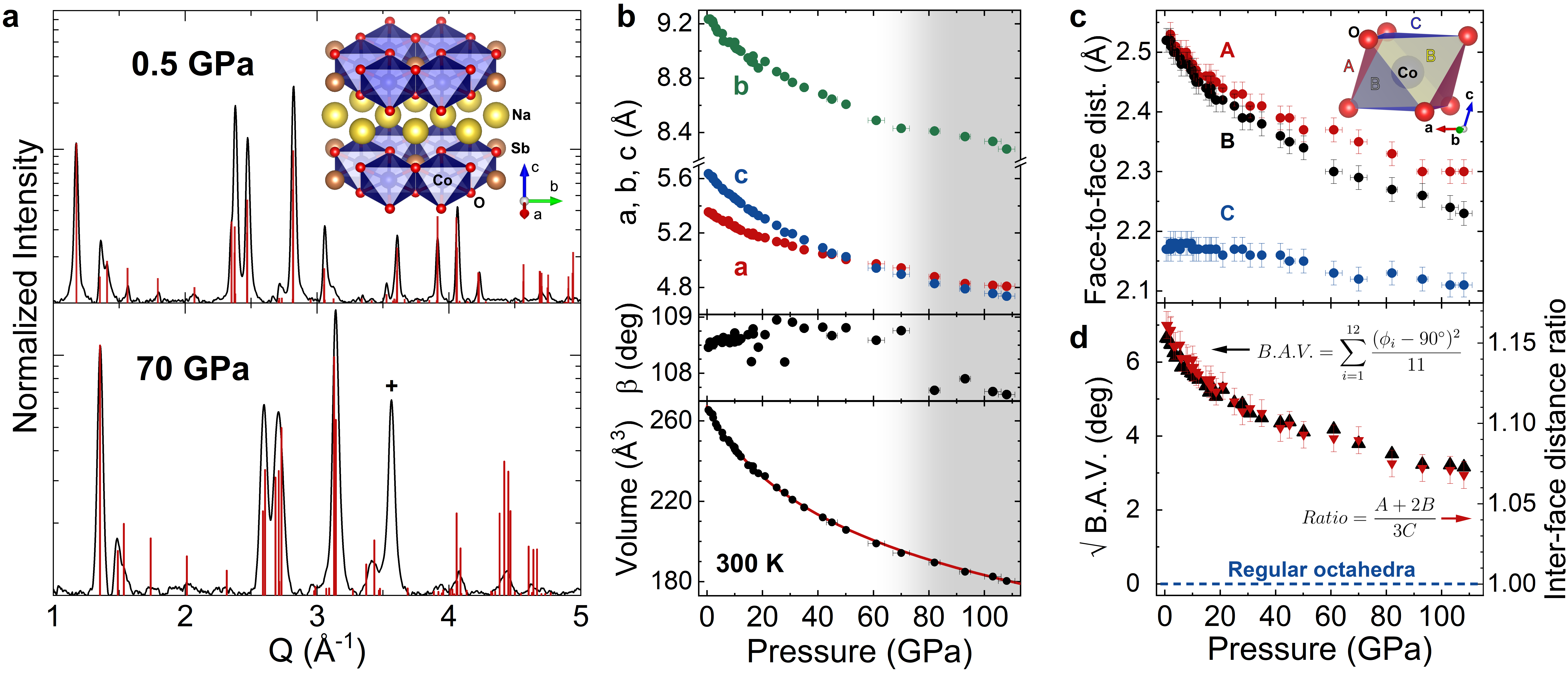}
    \caption{{\bf Evolution of crystal structure and trigonal distortion probed with x-ray diffraction.} \textbf{(a)} Diffraction patterns at 0.5 GPa and 70 GPa at room temperature, shown in log scale to enhance low intensity peaks. Symbol $+$ denotes a Bragg peak from the neon pressure transmitting medium. Red bars represent theoretically calculated Bragg peak positions (VESTA \cite{VESTA}) using the ambient pressure space group and atomic positions \cite{Viciu2007} and  experimental lattice parameters. Intensities are normalized to the most intense Bragg peak at $\sim 2.8~{\text{\AA}}^{-1}$. \textbf{(b)} Monoclinic lattice parameters $a$, $b$, $c$ and $\beta$ and unit cell volume $V$ (points) alongside fitted third-order Birch-Murnaghan equation of state (line). The shaded region above 70 GPa features lattice anomalies as a result of a spin transition. \textbf{(c)} Distances between O$_3$ facets in CoO$_6$ octahedra obtained from DFT+U calculations of relaxed atomic positions using experimental lattice parameters. Inset shows CoO$_6$ octahedron with pairs of facets labeled A, B and C, with monoclinic lattice vectors for reference. \textbf{(d)} Evolution of trigonal distortion with pressure quantified using bond angle variance and asymmetry in O$_3$ facet distances.}\label{fig:1}
\end{figure*}

X-ray powder diffraction data were collected as a function of pressure at selected temperatures in the 8 K to 300 K range. Full data sets and corresponding Le Bail fits are shown in Figs. S1 and S2 of the supplemental material. An expanded view (log scale) of 0.5 GPa and 70 GPa data is shown in Fig. \ref{fig:1}a. Le Bail fits using the known monoclinic structure at ambient pressure \cite{Viciu2007} provide a good description of the data over the entire pressure range. No new Bragg peaks are observed, indicating absence of  structural phase transitions, including dimerization. 

Figure \ref{fig:1}b shows the pressure dependence of lattice parameters. The pressure-volume relation, also shown in Fig. \ref{fig:1}b can be fitted by a single equation of state using either a 3rd-order Birch-Murnaghan model \cite{Birch1952} or Vinet model \cite{Vinet1986JPhysC,Vinet1987PRB}. The smooth P-V relation is also indicative of the absence of structural phase transitions. Both models yield consistent ambient pressure volume values of $V_0 = 266.2\pm0.3$ \AA$^3$. Fitted values for bulk modulus and its pressure derivative at 1 atm are $B_0 = 99\pm3$ ($97\pm3$) GPa and $B_0^{'} = 5.2\pm0.2$ ($5.5\pm0.2$) using the Birch-Murnaghan (Vinet) models, respectively. Despite the apparent absence of structural transitions, anomalies are observed in structural parameters above about 70 GPa as evidenced in the non-monotonic response of the $b$-axis and the monoclinic $\beta$ angle. As discussed below, a transition from HS to LS Co$^{2+}$ states is detected in the 60-70 GPa pressure range. The smaller atomic volume \cite{Shannon1976}, coupled with Jahn-Teller activity of LS divalent Co ions, is expected to manifest lattice anomalies. We note that Bragg peak broadening at the highest pressures and limited number of Bragg peaks accessible within the angular aperture of the diamond anvil cell (DAC) hinders detection of small peak splittings that would signal a lowering of symmetry, were it to take place. Forcing the monoclinic $\beta$ angle to its low-pressure value results in a small volume drop of $\sim$ 1.6\% around 70 GPa, albeit with a significantly worse fit quality (see Fig. S3). For comparison, PbCoO$_3$ shows no volume discontinuity at its HS to LS transition centered around 10 GPa \cite{Liu2020JACS}. CoO shows a 2.7\% volume change across a 90 GPa transition between rhombohedral phases with different densities, presumably connected with a spin transition \cite{Guo2002JPCM}, although XES shows the spin transition taking place at 140 GPa and only after laser heating \cite{Rueff2005JPCM}. 

Poor powder averaging as a result of the limited sample volume in the DAC, together with weak x-ray scattering power for low-Z oxygen ions, prevented reliable Rietveld refinement of atomic positions. High-pressure single crystal x-ray diffraction and neutron powder diffraction experiments were attempted. The presence of stacking faults and twin domains in our single crystals made it extremely challenging to reach a structural solution. A high background in neutron powder diffraction resulted in very few observable Bragg peaks preventing Rietveld refinements. Therefore, pressure-dependent experimental lattice parameters from Le Bail fits were used in DFT+U calculations of relaxed atomic positions to obtain the evolution of trigonal distortion with pressure. Figure \ref{fig:1}c shows the distances between pairs of opposite O$_3$ facets in the distorted CoO$_6$ octahedra. Figure \ref{fig:1}d quantifies the bond angle variance (BAV) of their twelve O-Co-O angles \cite{Robinson1971} alongside a normalized difference between in-plane and out-of-plane O$_3$ facet distances. At ambient pressure, CoO$_6$ octahedra are compressed along their trigonal axis, which is normal to the $ab$ honeycomb planes (close, but not aligned with, monoclinic $c$-axis). They become more regular under compression as the distances between A-A and B-B facets shrink faster than that between C-C facets, also leading to a reduction in BAV (regular octahedra have BAV = 0). Interestingly, the C-facets with surface normal oriented close to the $c$-axis are less affected by the anisotropic lattice compression of the layered structure (higher along $c$-axis) up to about 50 GPa, likely due to the ability of the spacer sodium layer between Co-Sb honeycomb layers to accommodate much of the $c$-axis compression up to this pressure. The trigonal distortion decreases continuously upon compression, roughly reaching half its ambient pressure value at the highest pressures.

\subsection*{Magnetism and local moments}
    
\begin{figure*}[ht]
    \centering
    \includegraphics[width=\textwidth]{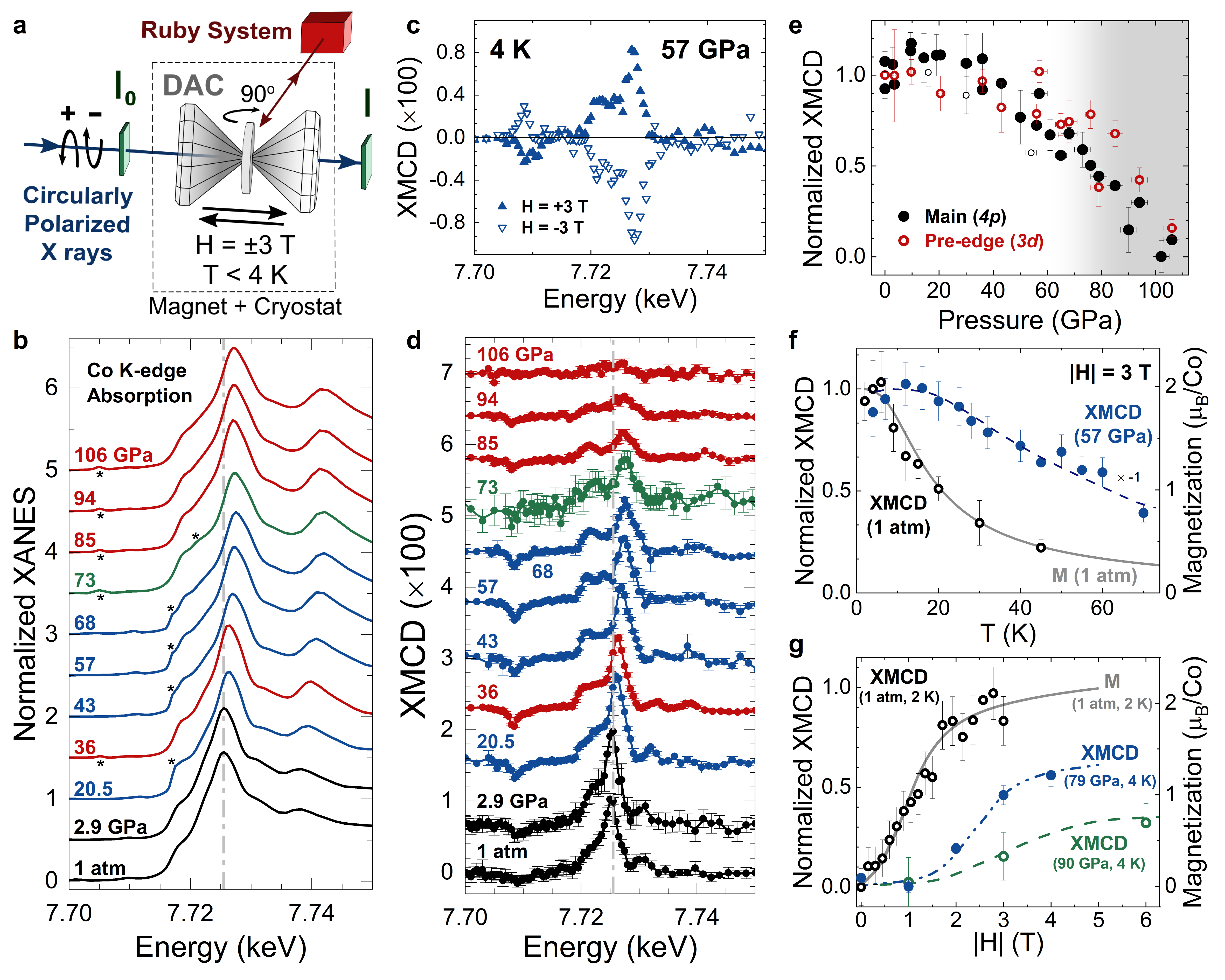}
    \caption{{\bf Evolution of electronic structure and spin correlations probed with x-ray absorption and magnetic circular dichroism.} \textbf{(a)} Experimental setup for XANES and XMCD experiments; see methods for details. \textbf{(b)} Selected Co K-edge normalized XANES and \textbf{(c,d)} corresponding XMCD as a function of pressure for a powder sample at $|H|=3$ T. Data in \textbf{(d)} were obtained by subtracting and halving XMCD data for opposite field directions. Ambient pressure data were collected at 2 K; data at other pressures at 4 K. Different colors denote different experimental runs; pressure values given in GPa units. Vertical dashed and dotted lines at 7.726 and 7.727 keV in \textbf{(b,d)} are guides to the eye. Star (*) symbols on XANES spectra indicate contamination signal from Bragg peaks in diamond anvils. \textbf{(e)} Integrated XMCD signals, normalized to 1 atm values, for the pre-edge (red open circles) and "white line" (closed black circles). The spin transition region, between about 60 and 90 GPa, is shown in gradient color bridging the white (HS) and gray (LS) regions. \textbf{(f)} Temperature dependence of integrated XMCD signal at 1 atm and 57 GPa. \textbf{(g)} Field dependence of integrated XMCD signal at 1 atm, 79 GPa, and 90 GPa. In \textbf{(f,g)}, ambient pressure magnetometry data (SQUID, solid lines) are displayed on the right axis. Normalized XMCD at 1 atm was scaled to match the magnetization data. Dashed lines are guides to the eye.}\label{fig:2}
\end{figure*}

The evolution of magnetism with pressure was probed with X-ray magnetic circular dichroism (XMCD) measurements at the Co K-edge on a powder sample. Figure \ref{fig:2}b shows the isotropic x-ray absorption spectra obtained by averaging over x-ray helicity, while Figs. \ref{fig:2}c,d show circular dichroic spectra. Measurements below 10 GPa were done at $T=2$ K, while measurements above this pressure were done at 4 K to prevent freezing of the pressurized He gas in compression and decompression membranes that actuate on the DAC piston for \textit{in-situ} pressure control. A 3 T field was applied to probe the evolution of magnetism within the in-field polarized phase. A weak pre-edge feature at 7.708 keV probes Co $3d$ states via quadrupolar $1s\rightarrow3d$ excitation while higher energy features, including the main ``white line" peak at $\sim7.726$ keV, correspond to dipolar $1s\rightarrow4p$ excitations. At ambient pressure, circular dichroic signals have peak intensity values of 0.2\% and 1\% at pre-edge and white line positions, respectively (Fig. S6). The quadrupolar pre-edge feature remains constant in energy upon compression, and so does the “on-edge” dipolar shoulder feature (some fine structure appearing on this on-edge feature in some of the spectra is an artifact of Bragg diffraction from diamond anvils, noted with $*$ symbols). 

The main white line peak and its related dichroic signal gradually shift to higher energy up to about $60-70$ GPa, for an overall shift of $\sim$ 1.8 eV (Fig. S5). The increased separation between white line and “on-edge” features results in the observable splitting of the dipolar dichroic signal. We attribute these energy shifts to increases in crystal field and/or hybridization involving the more extended Co $4p$ states. The energy shifts appear to slow down above $60-70$ GPa relative to expectation from volume contraction (Fig. S5). As discussed below, a spin transition is observed in this pressure range.

The rather similar line shape of the isotropic spectra over the entire pressure range is consistent with the absence of structural transitions, in line with results from XRD (the increase in amplitude of the high energy feature is due to stiffening of the lattice). The evolution of the integrated intensity of the dipolar and quadrupolar dichroic signals is shown in Fig. \ref{fig:2}e. Although the dipolar signal does not probe $3d$ magnetism directly, it is well established that the polarization of $4p$ states by the $3d-4p$ interaction enables use of $4p$ circular dichroism as a proxy of $3d$ magnetism. This is clearly seen by comparing the temperature and field dependence of conventional magnetometry data (Fig. S4) with that of the integrated intensity of dipolar XMCD signal, both obtained at 1 atm (solids lines and open circles in Figs. \ref{fig:2}f,g). While the quadrupolar dichroic pre-edge signal is harder to study systematically due to its smaller size, the evolution of its integrated intensity is in overall agreement with that of the dipolar feature (Fig. \ref{fig:2}e).

As seen in Fig. \ref{fig:2}e, the magnetization of the in-field ($H=3$ T) polarized phase measured at 4 K is rather constant up to about $50-60$ GPa, then decreases to vanish at about 100 GPa. The XMCD signal can be scaled to sample magnetization using ambient pressure magnetometry data (solid lines in Fig. \ref{fig:2}f,g). Temperature-dependent XMCD data collected at $P=57$ GPa, shown in Fig. \ref{fig:2}f (raw data in Fig. S6), point to enhanced ferromagnetic correlations at this pressure with sizable field-induced magnetization persisting to much higher temperatures than at ambient pressure. Despite the significant ferromagnetic correlation present in the field-induced polarized phase, it does not exhibit spontaneous ferromagnetic order. The enhancement in magnetic correlations can be crudely estimated using a simple Ising model with ferromagnetic nearest-neighbor interactions in the presence of a 3 T applied field, yielding approximate $J/k_B$ values $\sim$ 15 K and $\sim$ 50 K for 1 atm and 57 GPa, respectively (Fig. S7). Consequently, pressure appears to enhance the ferromagnetic interaction in the field-induced polarized phase, resulting in roughly a three-fold increase in the strength of ferromagnetic exchange at 57 GPa.

Field-dependent XMCD data at $T=4$ K for pressures above 60 GPa are shown in Fig. \ref{fig:2}g (raw data in Fig. S8). These data demonstrate an enhancement with pressure of the field required to induce polarization, and a strong reduction in saturation magnetization at 6 T. Furthermore, the magnetic susceptibility is progressively suppressed, ultimately vanishing at 100 GPa within the accuracy of our measurements. 

\begin{figure*}[ht]
    \centering
    \includegraphics[width=\textwidth]{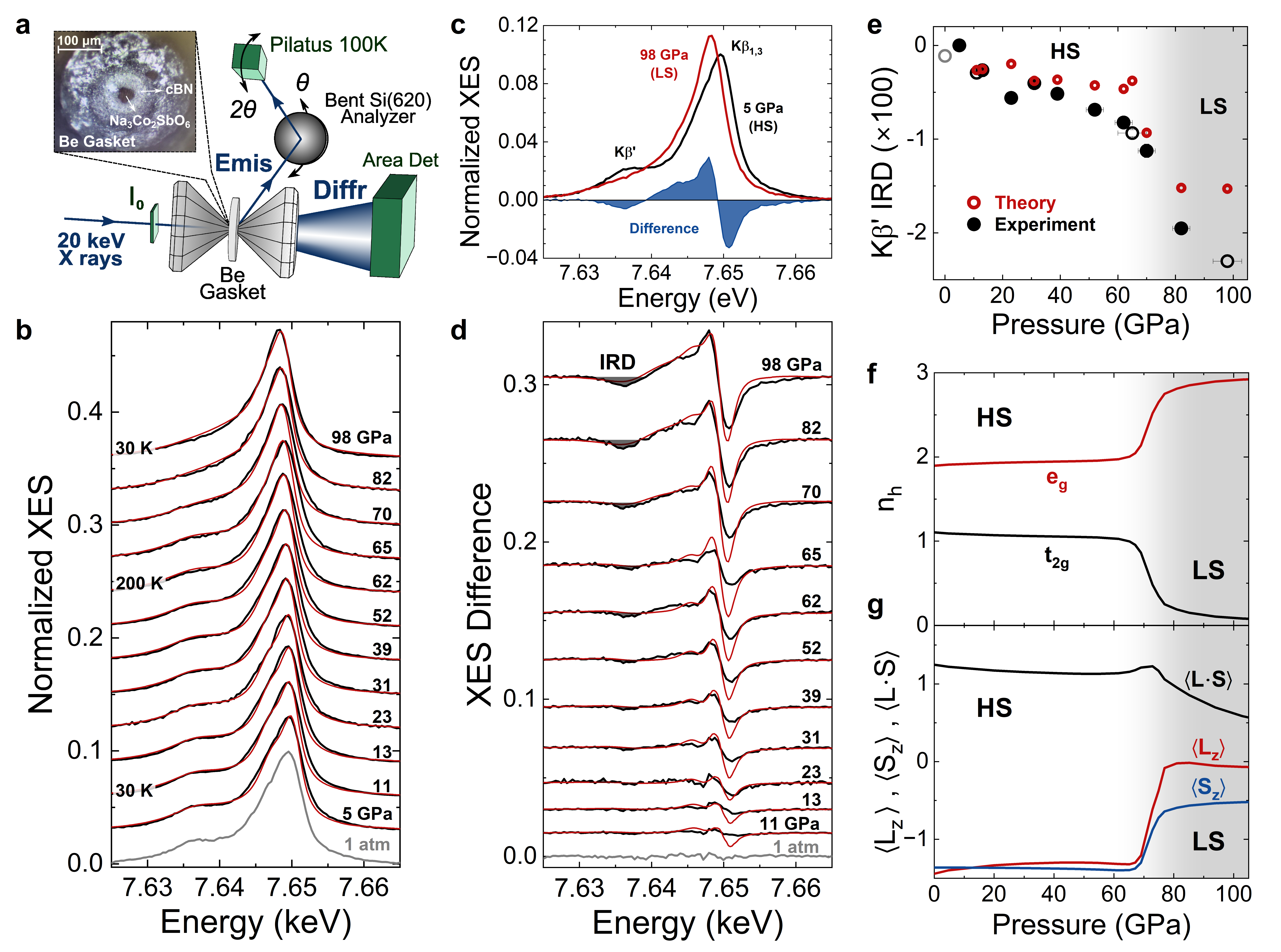}
    \caption{{\bf Evolution of $J_{\rm eff}=\frac{1}{2}$ pseudospins probed with x-ray emission spectroscopy.} \textbf{(a)} High-pressure XES experimental setup; see methods for details. \textbf{(b)} Area-normalized experimental (black) and theoretical (red lines) Co K$\beta$ emission spectra. Data were collected mostly at room temperature, with some spectra collected at lower temperatures as labeled. The 65 GPa data were collected during decompression. \textbf{(c)} Comparison between HS and LS spectra. \textbf{(d)} Difference in XES spectra relative to the 5 GPa spectra (lowest pressure in the DAC). Black lines are experiment; red lines theory. \textbf{(e)} XES integrated relative differences (IRD) as a function of pressure. Experimental (black circles) and theoretical (open red circles) data show a drop around 60-70 GPa signaling the beginning of a spin transition from HS (white region) to LS (gray region). Open black circles indicate experimental data collected at temperatures different from ambient, and the open gray circle represents 1 atm data collected outside of the DAC. \textbf{(f)} Calculated number of holes in Co $e_g$ and $t_{2g}$ states and \textbf{(g)} expectation values $\langle S_z\rangle$, $\langle L_z \rangle$ and $\langle \textbf{L}\cdot \textbf{S}\rangle$ for Co $3d$ electrons.}\label{fig:3}
\end{figure*}

To probe the evolution of $J_{\rm eff}=\frac{1}{2}$ pseudospins in Co$^{2+}$ ions, non-resonant Co $K\beta$ X-ray emission spectroscopy (XES) was employed. In the presence of a $3d$ spin moment, the two final states of $3p$ to $1s$ $K\beta$ emission (spin-up or spin-down hole in $3p$ orbital) are separated in energy by $3p-3d$ intra-atomic exchange resulting in a primary $K\beta_{1,3}$ line and a $K\beta^{\prime}$ satellite. The intensity of the satellite is proportional to the $3d$ spin moment, which also influences the energy separation between the emission lines \cite{Tsutsumi1959,Tsutsumi1968}. While a sensitive probe of local moment, XES is insensitive to spin correlations when measured without manipulation or analysis of X-ray polarization \cite{xes-mcd-footnote}. XES data as a function of pressure, collected primarily at 300 K but also at selected lower temperatures, are shown with black lines in Fig. \ref{fig:3}b. The presence of an intense $K\beta^{\prime}$ satellite at ambient pressure is as expected for the HS $S=\frac{3}{2}$ configuration of Co$^{2+}$ ions. XES difference spectra relative to 5 GPa (Fig. \ref{fig:3}d) highlight the changes in both primary line emission energy and intensity, as well as satellite intensity. The integrated relative difference (IRD), shown in Fig. \ref{fig:3}e, is the integral of the difference spectra over the energy range encompassing the satellite peak \cite{Li2019PRB,Yoo2005PRL,Mao2014AmMin} (difference spectra over the full energy range are shown in Fig. S9). The intensity of the $K\beta^{\prime}$ satellite decreases slowly up to about 60 GPa, likely a result of reduced intra-atomic $3p-3d$ overlap from increased delocalization of $3d$ valence electrons under compression. At pressures above 60 GPa a sudden acceleration in the loss of satellite intensity is observed, which we interpret to signal a spin transition from HS ($S=\frac{3}{2}$) to LS ($S=\frac{1}{2}$) state. Theoretical simulations of XES spectra, difference spectra, and IRD intensities using cluster calculations are shown with red lines and open circles alongside their experimental counterparts in b,d,e panels of Fig. \ref{fig:3}. A spin transition will take place when the HS state, which maximizes the total spin of $t_{2g}$ and $e_g$ orbitals (Hund’s first rule) is no longer energetically favorable in the presence of an increasing octahedral crystal field that raises the energy of doubly populated $e_g$ states under compression, leading to a more favorable $t_{2g}^6e_g^1$ LS configuration. Figure \ref{fig:3}f shows numerical calculations of hole occupancies, while Fig. \ref{fig:3}g shows the related evolution of expectation values $\langle S_z\rangle$, $\langle L_z \rangle$, and $\langle \textbf{L}\cdot \textbf{S}\rangle$ for Co $3d$ electrons. As expected, the orbital angular momentum nearly vanishes in the low spin state due to full occupation of $l_{\rm eff}=1$, $t_{2g}$ states. Density functional calculations find a HS-LS transition at about 70 GPa (Fig. S11), in agreement with the XES results. 

The decrease of magnetic susceptibility above 60 GPa is driven by the spin transition, as seen in the correlated suppression of XMCD and XES signals (Fig. S10a). The experimental transition width spans a change in reduced volume of about 5\% (Fig. S10b), which is similar to the 7-8\% span seen in PbCoO$_3$ \cite{Liu2020JACS}. The spin transition width has both extrinsic and intrinsic contributions. Pressure gradients, reported in the Methods section, contribute to the finite width due to coexistence of HS and LS states in different parts of the illuminated sample. Intrinsic contributions involve coupling between HS and LS states on the same Co site when their energies become proximate under pressure. The intrinsic width is determined by the energy scale of the mixing interaction(s). In the cluster calculations, a one-particle spin-orbit interaction of 66 meV leads to the finite width shown in Fig. \ref{fig:3}f. The similar transition widths seen in XMCD (T=4 K) and XES (T=300 K) data, as well as no significant difference in XES spectra for selected pressures at 30 K and 300 K (Fig. \ref{fig:3}e), indicate that 300 K is not sufficiently high to contribute significant broadening to the transition. While DFT+U calculations are bound to yield sharp spin transitions when the LS state becomes energetically favorable (Fig. S11b), extensions to dynamical mean field theory (DMFT) account for local dynamic electronic effects and lead to finite transition widths similar to experiment \cite{Kunes2008NatMat}. 


The LS $t_{2g}^6e_g^1$ state is expected to be Jahn-Teller active. Density functional theory shows enhanced splitting of Co-O distances across the spin transition, consistent with Jahn-Teller distortion (Fig. S11). A cooperative Jahn-Teller distortion may lead to a lowering of lattice symmetry in the LS state that could go undetected in our XRD data due to peak broadening at the highest pressures. We speculate that such cooperative Jahn-Teller distortion, which may entail ordering of $d_{x^2-y^2}$ and $d_{z^2}$ orbitals, could contribute to the apparent slow-down of energy shifts in XAS spectra at the spin transition (Fig. S5) as well as counteract the effect of a local volume contraction that is expected from the reduced atomic volume of LS divalent Co ions \cite{Shannon1976}. Single crystal XRD \cite{Ji2019Nature}, neutron diffraction \cite{Haberl2023SciRep}, and optical Raman scattering \cite{Baldini2011PRL} could provide additional insight into a possible lowering of crystal symmetry in the LS state.

\begin{figure}[ht]
\centering
\includegraphics[width= 3 in]{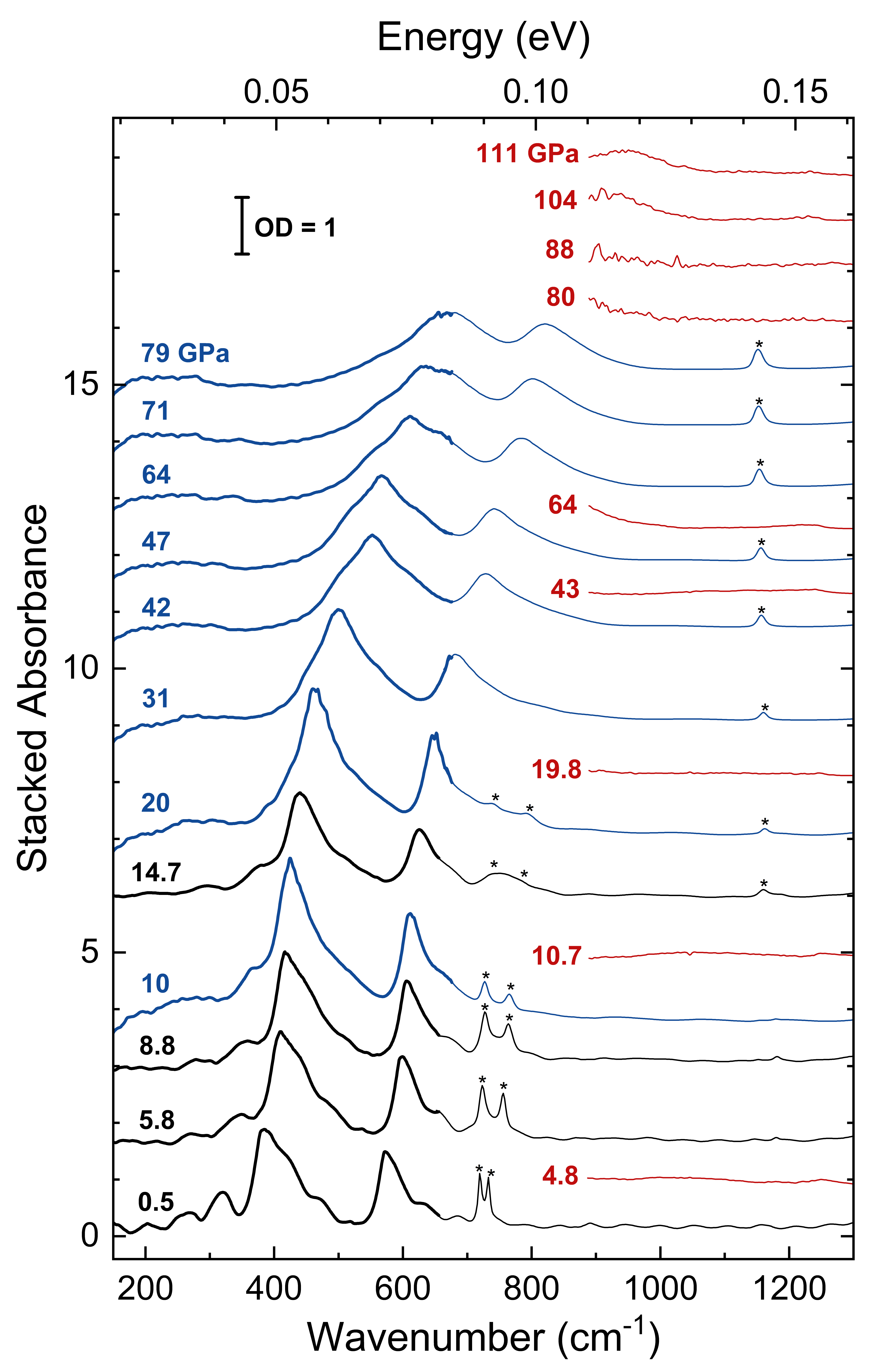}
    \caption{{\bf Absence of pressure-induced metallization probed with infrared spectroscopy.} Selected synchrotron far infrared (FIR, dark lines) and mid infrared (MIR, faint lines) absorbance spectra as a function of pressure. The scale bar on top denotes absorbance value of unity. Spectra are offseted in the vertical direction for clarity. Different colors represent different experimental runs. $*$ symbols denote excitations in the pressure transmitting medium (petrol jelly) in two of the runs. These peaks are absent on the third run where KBr medium was used instead.}\label{fig:4}
\end{figure}

\subsection*{Charge-transfer gap}

Na$_3$Co$_2$SbO$_6$ is a charge-transfer insulator in which the lowest charge excitation is dictated by the energy separation, $\Delta_{pd}$, between occupied O $2p$ and unoccupied Co $3d$ states. This charge-transfer gap, of order $2-3$ eV, as well as the larger $3d-3d$ ($U_{dd} \sim 5-8$ eV) excitation gap, are expected to stabilize the insulating state to pressures above the Mbar regime. An insulating state is required to stabilize the KQSL state where charge degrees of freedom are frozen out and the lowest energy physics is dictated by fractionalized spin excitations. Since suppression of spin correlations at highest pressure can also be a result of metallization, we carried out synchrotron infrared (IR) spectroscopy in both the far (FIR) and mid (MIR) ranges. IR absorbance data are shown in Fig. \ref{fig:4}. A metallic sample has perfect reflectivity for energies below the plasmon energy, $\hbar\omega_p$ (typical 5-10 eV). Absorbance data, computed as $\log(I_0/I)$ where $I_0$ is background intensity without sample and $I$ is transmitted intensity, is expected to diverge for $\omega\ll\omega_p$ as transmittance drops to zero. The FIR data shows phonon excitations that stiffen under pressure at a rate of $\sim 3.6(1)$ cm$^{-1}$/GPa (Fig. S12). Otherwise, the spectra reveal a high transmittance down to 20 meV with no signature of a Drude response from metallization. The MIR data reinforces this conclusion, not only by the lack of a Drude response to 111 GPa, but also from the absence of a closing charge-transfer gap excitation moving from the optical into the IR regime (multiple runs and full MIR data sets shown in Fig. S12). The MIR data set a lower limit of $\sim 1$ eV for the charge transfer gap at the highest pressure of 111 GPa.

\section*{Discussion}

The $J_{\rm eff}=\frac{1}{2}$ pseudospin wavefunction, and related superexchange interactions between pseudospins mediated by oxygen ions, depend on the size of the trigonal distortion of CoO$_6$ octahedra \cite{Liu2020PRL} which XRD shows is significantly reduced with pressure. Superexchange interactions are also influenced by the degree of covalency/mixing between Co $3d$ and O $2p$ orbitals, which is also modified upon lattice compression.  Based on the predictions of Liu {\it et al.} \cite{Liu2020PRL}, a reduction in trigonal crystal field should move Na$_3$Co$_2$SbO$_6$ towards the KQSL state. Since Kitaev exchange interactions in this cobaltate are expected to be ferromagnetic \cite{Liu2020PRL,Sano2018PRB,Songvilay2020PRB}, one may be tempted to assign the enhanced ferromagnetic correlations observed below 60 GPa, where the $J_{\rm eff}=\frac{1}{2}$ character of the pseudospin is preserved, to dominant Kitaev interactions. However, field-induced low-temperature magnetization near full moment values, which persists to rather high temperatures, is inconsistent with enhanced frustration. A likely explanation for the increased ferromagnetic correlations is pressure-induced enhancement of ferromagnetic $e_g-e_g$ isotropic superexchange interactions relative to the antiferromagnetic $t_{2g}-e_g$ superexchange, driven by increased covalency. Evidence for increased covalency is seen in the XES-IRD data which display a gradual reduction in intensity within the HS state below 60 GPa (Fig. \ref{fig:3}e). A near cancellation of isotropic ferro- and anti-ferromagnetic exchange interactions at ambient pressure is predicted to make this cobaltate proximate to a KQSL \cite{Liu2020PRL}, the sizable trigonal distortion preventing its full realization. Although the trigonal distortion is strongly reduced, a disruption in the balance of isotropic exchange interactions as a result of covalency will drive this cobaltate away from the KQSL limit at intermediate pressures.
    
We have assumed that the trigonal crystal field is dominated by the distortion of CoO$_6$ octahedra, i.e., by the oxygen ligands. Instead, if the trigonal field from Sb ions in Co-Sb planes dominates and contributes with an opposite sign, as has been previously proposed \cite{Liu2020PRL} but recently challenged \cite{Kim2022JPCM,Kim2024SciAdv}, removing the trigonal distortion of CoO$_6$ octahedra will increase the non-cubic crystal field and tend to stabilize magnetically ordered phases \cite{Liu2020PRL}. An increase in the ratio of Mott-Hubbard $U$ to charge transfer gap, $\Delta_{pd}$, under pressure can also contribute to an enhancement of non-Kitaev ferromagnetic correlations \cite{Liu2020PRL}.

The spin transition above 70 GPa manifests a strong suppression of $T=4$ K magnetic susceptibility, which vanishes at 100 GPa in 3 T applied field. The localized $3d$ orbitals protect this honeycomb lattice from a dimerization transition and therefore the mute susceptibility is not a result of formation of molecular orbitals with spin-singlet states, typical of dimerized $4d/5d$ honeycombs \cite{Takayama2019PRB,Clancy2018npj,Veenendaal2022,Takayama2022PRR}. 
Conventional antiferromagnetic ordering with gapped magnetic excitations can also lead to mute susceptibility and a similar field dependence to that measured at 90 GPa. However, first-neighbor exchange interactions between $e_g$ spins in LS Co$^{2+}$ ions are expected to be ferromagnetic \cite{Kanamori1959,Goodenough}, making this scenario less likely. The spin transition quenches the orbital angular momentum (filled $t_{2g}$ states) destroying the spin-orbital $J_{\rm eff}=\frac{1}{2}$ pseudospin wavefunction required to map bond-directional exchange interactions into Kitaev’s model. The emergence of strongly reduced, $S=1/2$ moments in a 2D honeycomb lattice of LS Co$^{2+}$ ions can lead to non-classical magnetic ground states as a result of strong quantum fluctuations. A $J_1-J_3$ ferro-antiferromagnetic $XXZ$ quantum $S=\frac{1}{2}$ model on the honeycomb lattice \cite{Jiang2023PRB}, where $J_1<0$ is ferromagnetic as expected for nearest neighbor interactions between $e_g$ spins, reveals a non-classical Ising-$z$ phase near the $XY$ limit and $J_3/J_1\sim -0.35$. It features compensated spins perpendicular to the honeycomb plane despite dominant in-plane exchange interactions, an example of quantum order by disorder. A study of the isotropic $J_1-J_2$ quantum $S=\frac{1}{2}$ model on the honeycomb lattice with ferromagnetic $J_1<0$ yields a gapped quantum spin liquid phase with short-ranged dimer-dimer correlations at $J_2/J_1\sim -0.25$ \cite{Fouet-J1J2}. In the absence of metallization, as clearly established by IR spectroscopy, exchange correlations between $S=\frac{1}{2}$ moments are expected to stabilize magnetic order at low temperature unless quantum fluctuations are at play. The muted susceptibility alongside its field-dependence measured in the LS state (Figs. \ref{fig:2}e,g) is consistent with emergent quantum paramagnetism as a result of frustrated exchange interactions between localized $S=\frac{1}{2}$ spins as found in $J_1-J_2-J_3$ quantum models in  honeycomb lattices \cite{Jiang2023PRB,Fouet-J1J2}. The measurement uncertainties at low applied fields prevent asserting whether such paramagnetic state is gapped or gapless. Further technical developments are needed to probe spin correlations and excitations into the Mbar range, e.g., with inelastic neutron scattering \cite{Zalyzniak-INS,Pajerowski-INS}, optical Raman scattering \cite{Li2021PRB}, or NMR techniques \cite{Haase-NMR} in order to provide additional insight into the complex quantum magnetism of highly compressed Na$_3$Co$_2$SbO$_6$.

\section*{Methods}

\subsection*{Sample synthesis}
Polycrystalline samples were synthesized at Argonne National Laboratory following the procedures described in Ref. \cite{Veenendaal2023}. Laboratory powder XRD confirmed the single phase nature of the samples. Magnetic susceptibility measurements at ambient pressure confirmed antiferromagnetic ordering below T$_{\rm N}$=8 K (see Fig. S4). Single crystalline samples were prepared as described in Ref. \cite{Yan2019PRM}, and displayed a N\'eel temperature T$_{\rm N}\sim$ 5 K. The reduced ordering temperature relative to powder samples was attributed to significant presence of stacking faults, promoted by the vapor transport growth method \cite{Yan2019PRM}. Improvements in single crystal growth using the flux method are now yielding T$_{\rm N}$ values within $\sim 1$ K of those in powder samples \cite{Gu2024PRB}. 
   
\subsection*{X-ray diffraction}
Diffraction measurements were performed at the Advanced Photon Source (APS), Argonne National Laboratory, using the High Pressure Collaborative Access Team (HPCAT) beamlines 16-BM-D and 16-ID-D \cite{HPCAT}. X-ray energies were 30 and 29.2 keV, respectively. Measurements were performed at 300 K, 240 K, 15 K and 8 K under variable pressures up to 108 GPa. Complete data sets are included in the supplemental information. Symmetric diamond anvil cells (DACs) were mounted with either 100 $\mu$m, 180$\mu$m or 300 $\mu$m-culet diamond anvils depending on the pressure range aimed for the run. Re gaskets were indented to a thickness of about $1/5$ of the culet diameter, and a hole was laser drilled in the center of the indentation with a diameter between $1/3$ and $1/2$ of the culet size \cite{Dunstan1989}. DACs with 100 $\mu$m-culet anvils were gas loaded \cite{Rivers2008} with He and used ruby luminescence \cite{Barnett1973,Chijioke2005} together with Raman spectroscopy of diamond \cite{Hanfland1985,Akahama2010} for pressure calibration. DACs with 180 $\mu$m and 300 $\mu$m anvils had either helium or neon as pressure medium, with both ruby and gold as pressure markers. Loose powder with grain size of about 1 $\mu$m was loaded without filling the entire sample chamber to achieve a reasonable sample-medium ratio for better hydrostaticity. Beam dimensions in all runs were around 5x5 $\mu$m$^2$ (FWHM) but the full width at 1\% of maximum intensity was around 25x25 $\mu$m$^2$. Diffraction patterns were integrated using Dioptas \cite{Dioptas} and Le Bail fitted using Jana2020 \cite{JANA} in order to get lattice parameters as a function of pressure. Volume at 1 atm, bulk modulus and its pressure derivative were obtained using the EosFit7 software \cite{EOSfit}.
  
\subsection*{X-ray magnetic circular dichroism}
Co K-edge XMCD measurements were performed at APS beamline 4-ID-D \cite{4IDD}, using loose powder loaded in CuBe DACs with ruby pressure marker\cite{Barnett1973,Chijioke2005} in neon medium \cite{Rivers2008}. Fully perforated, partially perforated and full anvils were used for experiments up to 106 GPa. 100, 180 and 300 $\mu$m-culet diamonds were employed in 4 different experimental runs, in which Re gaskets were indented to a thickness of about $1/5$ of the culet diameter, and a hole was laser drilled in the center of the indentation with a diameter between $1/3$ and $1/2$ of the culet size \cite{Dunstan1989}. Compression and decompression gas membranes were attached to the DAC body to allow for pressure stability during cool downs, and for reversibility studies. Pressure was calibrated before and after measurements at each pressure point, via a ruby system composed of a 473 nm laser, spectrometer and focusing optics. This system inserts into the reentrant room temperature bore between the split, longitudinal field magnet coils of a 6.5 Tesla cryomagnet, giving optical access to ruby fluorescence after rotating the DAC by $\sim90^{\circ}$ inside the magnet. Sample is cooled in $^4$He vapor. K-B mirrors were employed in the 3 last runs, which significantly improved signal-to-noise ratio in comparison with the first run (300 $\mu$m culets). The small sample volume used with 100 $\mu$m anvils (initial gasket indentation of about 20 $\mu$m) was the main source of noise for the highest pressures (reduced absorption edge jump). A combination of helicity switching of circularly polarized x-rays and magnetic field switching was used to obtain clean XMCD spectra and mitigate the contribution of artifacts. On average, each pressure point took from 8 to 10 hours to complete. Experiments were run at $|$H$|$=3 T and 2 K for pressures below 9.6 GPa. For all other pressure points experiments were run at 4 K to avoid freezing of Helium in the compression and decompression membranes. Pressure gradients were about ±3 GPa at the highest pressure.

\subsection*{X-ray emission spectroscopy}
High pressure Co K$\beta$ XES experiments were performed at HPCAT's beamline 16-ID-D, using 20 keV x-rays and a symmetric DAC in a diamond-in, gasket-out geometry. The DAC was placed inside a cryostat, pressure was applied using a gas membrane and XES data were collected through a side window after diffraction from a bent Si(620) analyzer. Single-beveled 100 $\mu$m-culet diamonds were used to indent a Be gasket to a thickness of about 20 $\mu$m. A hole of 98 $\mu$m was laser drilled on the indentation and the region was filled with a properly cured 10:1 cBN:epoxy (EPO-TEK 353ND) mixture \cite{Lin2003,Lin2008}. After packing, the cBN mixture formed a robust insert that was laser drilled to create the sample space, a 30 $\mu$m-diameter hole. Fine ($<$ 1 $\mu$m grain size) loose powder was loaded and silicone oil was used as pressure medium. X-ray diffraction was collected through the downstream cryostat window in forward scattering geometry using a Pilatus 100K detector. Pressure was calibrated using the (001) peak and the equation of state shown in Fig. \ref{fig:1}b. All XES data were analyzed using the integrated relative difference (IRD) method \cite{Li2019PRB,Mao2014AmMin} with focus on the pressure-dependent behavior of the satellite peak at $\sim7.636$ keV (see supplemental information). Pressure gradients were about ±5 GPa at the highest pressure.

\subsection*{Infrared absorption spectroscopy}
Infrared absorption spectra were collected at the frontier synchrotron infrared spectroscopy (FIS) beamline, 22-IR-1, Brookhaven National Laboratory (BNL). Absorbance, defined as $A=$ log$(I_0/I)$ where $I_0$ is the reference spectrum or background intensity and $I$ is the intensity of infrared light transmitted through the sample, was measured using single crystal samples \cite{Yan2019PRM}. High-pressure absorbance was measured in multiple runs using different DAC loadings covering the far infrared (FIR, $<650$ cm$^{-1}$) and mid infrared (MIR, $650-8000$ cm$^{-1}$) regions. Petrol jelly was used as pressure transmitting medium for the FIR region, and KBr for the MIR measurements. Due to different experimental setups, $I_0$ in the FIR region is measured by transmitting through the diamond anvils only (taken once at 1 atm before the sample is loaded), while in the MIR region it is measured through both diamonds and KBr at each pressure point. 
    
\subsection*{Density functional theory}
We adopted Density Functional Theory plus Hubbard U (DFT+U) \cite{Liechtenstein1995} based on the projected-augmented wave (PAW) method \cite{Bloch1994} as implemented in the Vienna ab initio simulation package (VASP) \cite{Kresse1996,Kresse1999}. The exchange-correlation energy functional was treated using generalized gradient approximation (GGA) by adopting the Perdew-Burke-Ernzerhof (PBE) functional \cite{Perdew1997}. We used the cutoff energy for the plane-wave basis as 600 eV, and the Gamma-centered $8\times4\times8$ k-point. To treat the correlation effect of Co $3d$ orbitals, we impose the Hubbard $U$ and the Hund’s coupling $J$ within DFT+U with $U = 5$ eV and $J = 0.8$ eV. Calculations were done both by relaxing atomic positions at different pressures while keeping lattice parameters fixed to their experimental values, as well as relaxing lattice parameters and atomic positions while keeping the unit cell volumes fixed to their experimental values. The Hellmann-Feynman force on each atom was set to be smaller than 0.01 eV/\AA\,for convergence.

\subsection*{Cluster calculations}
In order to describe the $K\beta$ spectral lineshapes and the ground-state expectation values, calculations were performed on a divalent cobalt ion. To describe the local electronic structure on cobalt, the same approach is followed as in Ref. \cite{Veenendaal2023,Veenendaal2015Book} but a CoO$_6$ is used in this case instead of a cobalt ion. The Hamiltonian includes the Coulomb and spin-orbit interactions. The effect of the ligands is included by an effective crystal field, $10Dq$. The trigonal crystal field interaction was not included as it is much weaker than $10Dq$ and Coulomb interactions driving the spin transition. The $K\beta$ spectra are obtained by calculating the $3p\rightarrow 1s$ radiative decay \cite{Wang1997PRB}. The crystal field values were converted to pressure using the following conditions. The $10Dq$ value at ambient pressure is 1.1 eV, which is obtained from a detailed fit of the $L$-edge XAS \cite{Veenendaal2023}. The spin crossover occurs at $10Dq\cong 2.27$ eV. The change in Co-O distances $d$ should follow the behavior of the lattice parameters. The crystal field depends on the lattice parameters as $(d_0/d)^a$ where $d_0$ is the distance at ambient pressure \cite{Harrison}. In order to satisfy these conditions, a value of $a\cong5$ is needed. This is close to the expected power for a change in the crystal field due to a change in metal-ligand distance.

\section*{Acknowledgements}

The authors thank Hide Takagi and Janice Musfeldt for helpful discussions. Work at the Advanced Photon Source (APS) and Materials Science Division of Argonne National Laboratory (ANL) was supported by the U.S. DOE Office of Science, Office of Basic Energy Sciences, under Contract No. DE-AC02-06CH11357. Work at UIC was supported by the U.S. National Science Foundation (NSF, DMR-210488), and the U.S. Department of Energy-National Nuclear Security Administration (DOE-NNSA) through the Chicago/DOE Alliance Center (DE-NA0004153), and the DOE Office of Science (DE-SC0020340). Portions of this work were performed at HPCAT (Sector 16) of APS at ANL. HPCAT operations are supported by DOE-NNSA’s Office of Experimental Sciences. Helium and neon pressure media were loaded at GeoSoilEnviroCARS (The University of Chicago, Sector 13), APS, ANL. GeoSoilEnviroCARS is supported by NSF - Earth Sciences (EAR-1634415) and DOE - GeoSciences (DE-FG02-94ER14466). Work at ORNL was supported by the U.S. DOE, Office of Science, Basic Energy Sciences, Materials Sciences and Engineering Division. R.T. was supported by the São Paulo Research Foundation (FAPESP, 2019/10401-9 and 2022/03539-7). H.P. acknowledges the support by the Materials Sciences and Engineering Division, Basic Energy Sciences, Office of Science, US DOE. We acknowledge the computing resources provided on Bebop, a high-performance computing cluster operated by the Laboratory Computing Resource Center at ANL. This research used the 22-IR-1 beamline (FIS) of the National Synchrotron Light Source II, a U.S. Department of Energy (DOE) Office of Science User Facility operated for the DOE Office of Science by Brookhaven National Laboratory under Contract No. DE-SC0012704. Work at UAB was supported by NSF CAREER Award No. DMR-2045760.


%

\end{document}


\def\nacosbo{$\mathrm{Na_3Co_2SbO_6}$}
\title{Supplemental Material for ``Pressure tuning of Kitaev spin liquid candidate Na$_3$Co$_2$SbO$_6$"}

\author{E. H. T. Poldi}
    \affiliation{Department of Physics, University of Illinois at Chicago, Chicago, Illinois 60607, USA}
    \affiliation{Advanced Photon Source, Argonne National Laboratory, Argonne, Illinois 60439, USA}
\author{R. Tartaglia}
    \affiliation{Institute of Physics Gleb Wataghin, University of Campinas, Campinas, Sao Paulo 13083-859, Brazil}
    \affiliation{Advanced Photon Source, Argonne National Laboratory, Argonne, Illinois 60439, USA}
\author{G. Fabbris}
    \affiliation{Advanced Photon Source, Argonne National Laboratory, Argonne, Illinois 60439, USA}
\author{N. Nguyen}
    \affiliation{Department of Physics, University of Illinois at Chicago, Chicago, Illinois 60607, USA}
\author{H. Park}
    \affiliation{Department of Physics, University of Illinois at Chicago, Chicago, Illinois 60607, USA}
\author{Z. Liu}
    \affiliation{Department of Physics, University of Illinois at Chicago, Chicago, Illinois 60607, USA}
\author{M. van Veenendaal}
    \affiliation{Department of Physics, Northern Illinois University, DeKalb, Illinois 60115, USA}
\author{R. Kumar}
    \affiliation{Department of Physics, University of Illinois at Chicago, Chicago, Illinois 60607, USA}
\author{G. Jose}
    \affiliation{Department of Physics, University of Alabama at Birmingham, Birmingham, Alabama 35294, USA}
\author{S. Samanta}
    \affiliation{Department of Physics, University of Alabama at Birmingham, Birmingham, Alabama 35294, USA}
\author{W. Bi}
    \affiliation{Department of Physics, University of Alabama at Birmingham, Birmingham, Alabama 35294, USA}
\author{Y. Xiao}
    \affiliation{HPCAT, Advanced Photon Source, Argonne National Laboratory, Argonne, Illinois 60439, USA}
\author{D. Popov}
    \affiliation{HPCAT, Advanced Photon Source, Argonne National Laboratory, Argonne, Illinois 60439, USA}
\author{Y. Wu}
    \affiliation{High Flux Isotope Reactor, Oak Ridge National Laboratory, Oak Ridge, Tennessee 37831, USA}
\author{J.-W. Kim}
    \affiliation{Advanced Photon Source, Argonne National Laboratory, Argonne, Illinois 60439, USA}
\author{H. Zheng}
    \affiliation{Materials Science Division, Argonne National Laboratory, Argonne, Illinois 60439, USA}
\author{J. Yan}
    \affiliation{Materials Science and Technology Division, Oak Ridge National Laboratory, Oak Ridge, Tennessee 37831, USA}
\author{J. Mitchell}
    \affiliation{Materials Science Division, Argonne National Laboratory, Argonne, Illinois 60439, USA}
\author{R. J. Hemley}
    \affiliation{Departments of Physics, Chemistry and Earth and Environmental Sciences, University of Illinois at Chicago, Chicago, Illinois 60607, USA}
\author{D. Haskel}
    \email{haskel@anl.gov}
    \affiliation{Advanced Photon Source, Argonne National Laboratory, Argonne, Illinois 60439, USA}

\maketitle
\setcounter{figure}{0}
\renewcommand{\figurename}{Figure}
\renewcommand{\thefigure}{S\arabic{figure}}


\section*{X-ray diffraction}
    
\begin{figure}[ht]
    \centering
        \includegraphics[width=6.4 in]{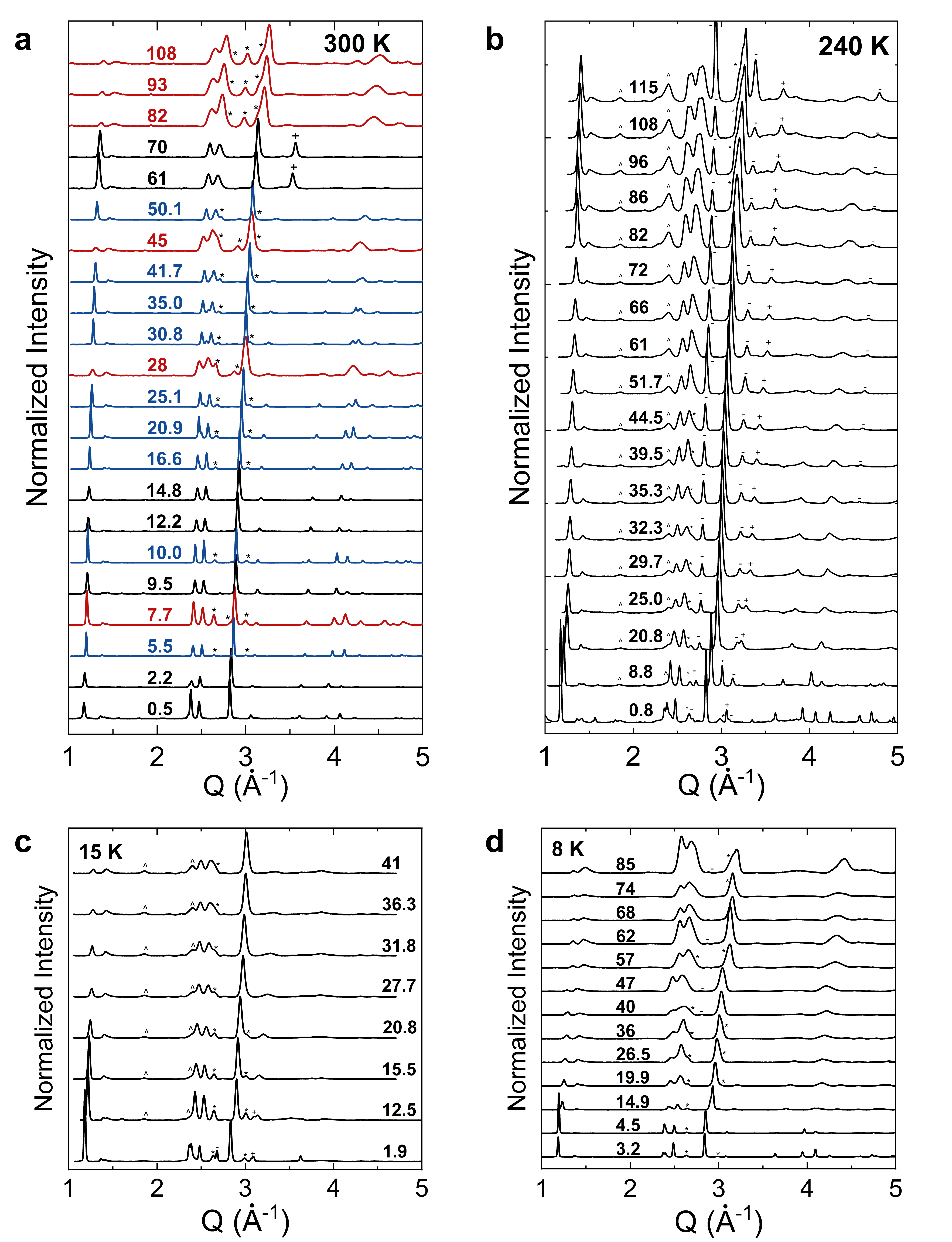}
        \caption{{\bf Pressure-dependent diffraction patterns at selected temperatures.} XRD patterns as function of pressure (in GPa) at \textbf{a} 300 K, \textbf{b} 240 K, \textbf{c} 15 K, and \textbf{d} 8 K. Different colors correspond to different experimental runs. Symbols $-$, $*$, $+$ and $\hat{}$ denote Au, Re, Ne and cryostat window peaks, respectively. All data were normalized to unity at the most intense Bragg peak, located at $\sim2.8$ \AA$^{-1}$ at the lowest pressures 0.5, 0.8, 1.9 and 3.2 GPa for each of the temperatures.}\label{fig:S1}
    \end{figure}

    Figure \ref{fig:S1} shows diffractograms at various temperatures and pressures. The data are normalized to unity at the most intense Bragg peak, located at $\sim2.8$ \AA$^{-1}$ at the lowest pressures 0.5, 0.8, 1.9 and 3.2 GPa for each of the temperatures.  Different colors represent different experiment runs: black and blue patterns were obtained with 300 $\mu$m-culet DACs gas loaded with Ne and He respectively; while red  used a 100 $\mu$m DAC loaded with He. Some sets of experiments present diffraction patterns with the lowest order (001) peak suppressed at higher pressures. This observation is not consistent throughout all sets of experiments, occurring only in some runs where smaller diamond anvil culets were employed (100 $\mu$m). Single crystals grow in  platelet-like shape with $c$-axis normal to the surface \cite{Yan2019PRM}. Powder grains may preserve the same morphology which can lead to powder texturing. The larger sample-to-medium ratio in DACs with smaller diamond culets will accentuate this effect. While texturing will also affect intensity of higher order (00l) peaks, those have lower intensities and also overlap with multiple other Bragg peaks making it difficult to observe texturing effects at these peaks. Diffraction peaks from the cryostat window display increased relative intensity at highest pressures due to a weakening of the Bragg peak used as reference for normalization.
    
\begin{figure}[ht]
    \centering
        \includegraphics[width=6.5 in]{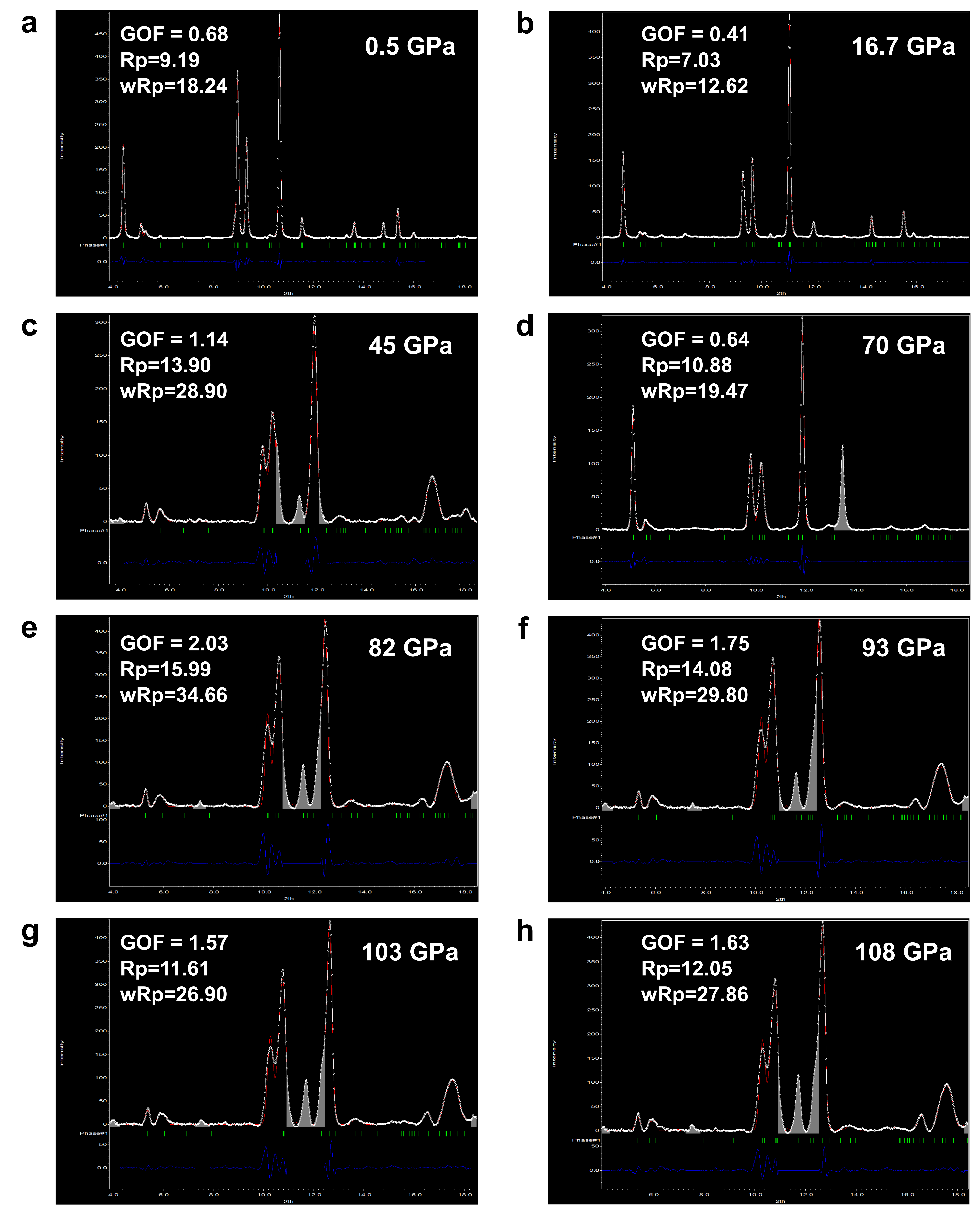}
        \caption{\textbf{Le Bail fits for selected diffraction patterns at 300 K.} The monoclinic angle $\beta$ was free to vary during these refinements. Pressures and quality of fit values \cite{Toby2006} are indicated in each panel. Shaded (gray) areas are regions excluded from Le Bail fittings due to the contamination of spurious diffraction peaks from W, Ne, or diamond anvil carbide seats. Vertical tick marks are refined Bragg peak positions and misfit is shown on the bottom. CIF files of the refined structures can be downloaded at https://doi.org/10.5281/zenodo.14052640.}\label{fig:S2}
    \end{figure}

    Figure \ref{fig:S2} shows Le Bail fits of XRD data (300 K) for selected pressures together with goodness of fit parameters \cite{Toby2006}. Although the XRD analysis is limited due to preferred orientation and presence of Au, Re and/or Ne diffraction peaks from pressure marker and gasket, Le Bail fits using the known monoclinic structure at ambient pressure\cite{Viciu2007,Yan2019PRM,Stratan2019} provide a good description of the data over the entire pressure range with no structural phase transitions, including dimerization, detected. Volume at 1 atm, bulk modulus and its pressure derivative were obtained using the EosFit7 software \cite{EOSfit}.
    
\begin{figure}[ht]
    \centering
        \includegraphics[width=5 in]{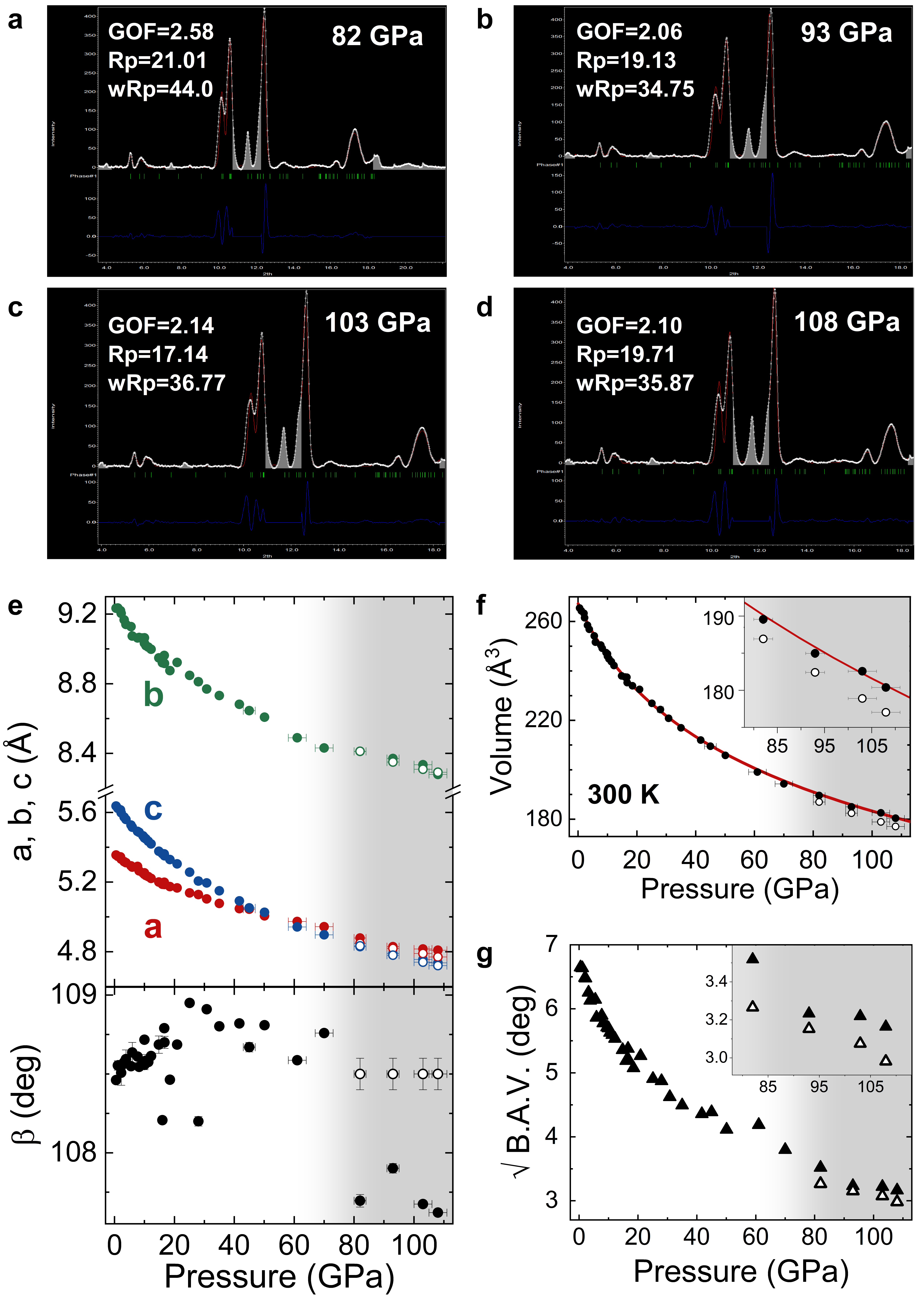}
        \caption{{\bf Impact of fixing monoclinic $\beta$ lattice parameter to its low-pressure value.} \textbf{a-d} Le bail fittings at 82, 93, 103 and 108 GPa for $\beta$ fixed to 108.5$^{\circ}$ during refinement. Shaded (gray) areas are regions excluded from Le Bail fittings due to the contamination of spurious diffraction peaks from W, Ne, or diamond anvil carbide seats. Vertical tick marks are refined Bragg peak positions and misfit is shown on the bottom. \textbf{e} Unit cell lattice parameters, \textbf{f} volume, and \textbf{g} square root of the bond angle variance (BAV) for CoO$_6$-octahedra as a function of pressure. Closed (colored) symbols are derived from Le Bail fits where monoclinic $\beta$ angle was a fitting parameter, while white symbols correspond to fits where $\beta$ was fixed to its low-pressure value. The spin transition region, between about 60 and 90 GPa, is shown in gradient color bridging the white (HS) and gray (LS) regions. CIF files of the refined structures can be downloaded at https://doi.org/10.5281/zenodo.14052640.}\label{fig:S3}
    \end{figure}

    The anomalies seen in the monoclinic $\beta$ angle across the spin transition, together with the absence of a measurable volume change at this transition, led us to evaluate the effect of forcing $\beta$ to its low pressure value. Le Bail fits of 300 K XRD data carried out using a fixed $\beta=108.5^{\circ}$ value for data at pressures above 70 GPa, i.e., in the low-spin state, are shown in Fig. \ref{fig:S3}. The goodness of fit parameters are worse than those obtained with variable $\beta$ (Fig. \ref{fig:S2}), with the latter providing a better description of the data. Nevertheless, derived lattice parameters, volume, and bond angle variance of CoO$_6$ octahedra obtained with a fixed $\beta$ value are shown in Fig. \ref{fig:S3} alongside the results obtained with variable $\beta$. A small volume discontinuity, of about 1.6\%, is seen when $\beta$ is fixed to its ambient pressure value. Although our best fits point to the absence of a volume discontinuity at the spin transition, limitations in the XRD data due to broadening at highest pressures call for additional investigation of structural effects at the spin transition. Single crystal XRD experiments were attempted, but stacking faults and twin domains present in our crystals prevented us from reaching a structural solution. Table \ref{tab:lat-param} shows refined lattice parameters above 70 GPa for the variable and fixed $\beta$ scenarios.
        
    \begin{table}
        \centering
        \begin{tabular}{|c|c|c|c|c|c|c|c|}
            \hline
            \multirow{2}{*}{Pressure (GPa)} & \multicolumn{4}{c|}{Variable $\beta$} & \multicolumn{3}{c|}{Fixed $\beta$} \\\cline{2-8}
             & a (\AA) & b (\AA) & c (\AA) & $\beta$ (deg) & a (\AA) & b (\AA) & c (\AA) \\\hline
            82 & 4.879(2) & 8.412(4) & 4.829(3) & 107.70(4) & 4.848(3) & 8.412(8) & 4.834(3) \\
            93 & 4.829(2) & 8.371(5) & 4.791(2) & 107.90(3) & 4.819(2) & 8.351(6) & 4.778(3) \\
            103 & 4.816(2) & 8.335(3) & 4.756(1) & 107.68(2) & 4.789(2) & 8.309(6) & 4.740(3) \\
            108 & 4.809(2) & 8.279(3) & 4.736(1) & 107.62(2) & 4.770(2) & 8.292(5) & 4.717(3) \\\hline
        \end{tabular}
        \caption{\textbf{Comparison of refined lattice parameters for high-pressure structures.} The monoclinic angle $\beta$ was either refined (allowed to vary) on Jana2020 \cite{JANA}, or set to its near-ambient value (fixed at 108.5 deg).}
        \label{tab:lat-param}
    \end{table}

\section*{Magnetization, X-ray Absorption, and X-ray Magnetic Circular Dichroism}
    
\begin{figure}[ht]
    \centering
        \includegraphics[width=6.5 in]{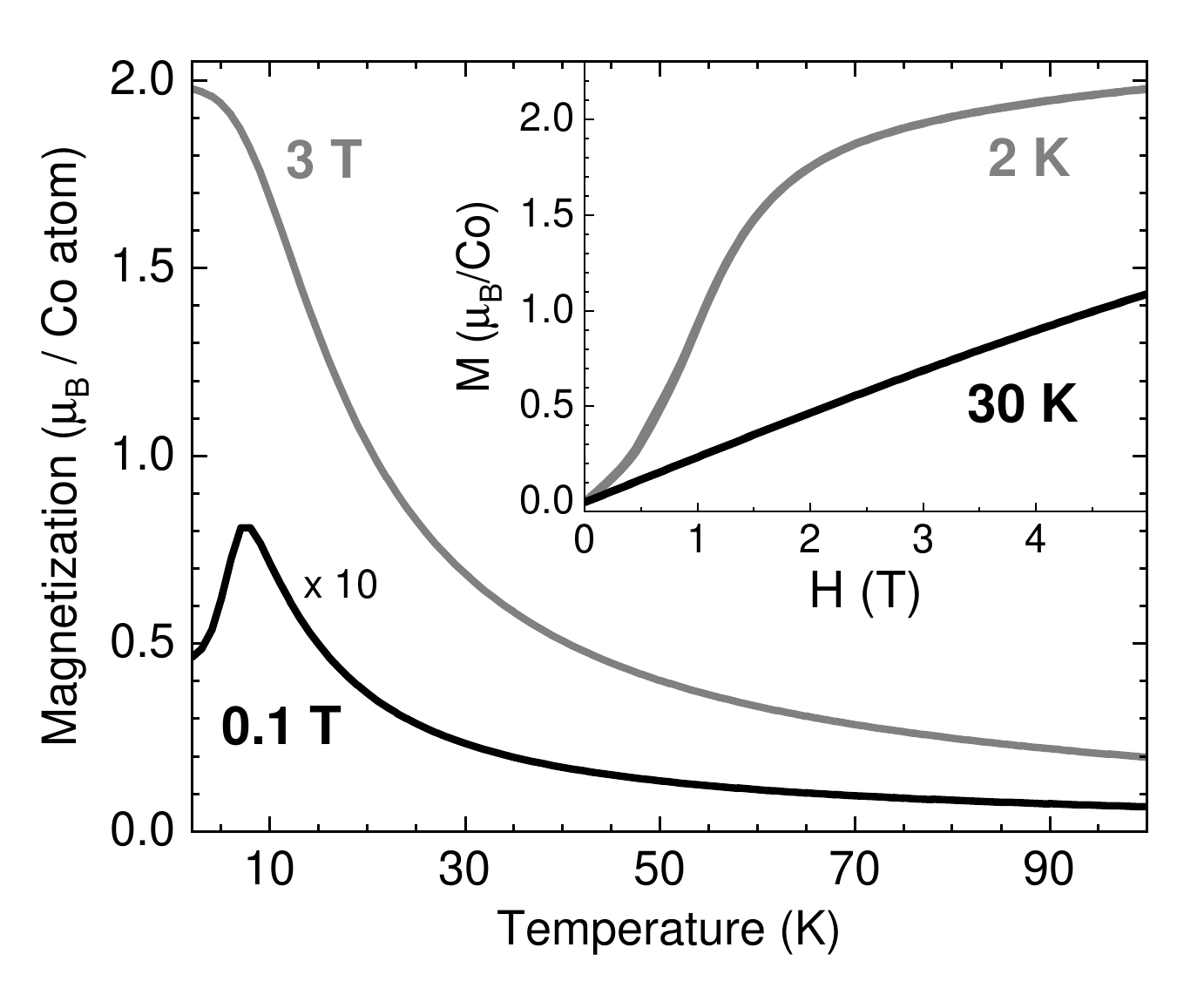}
        \caption{{\bf Magnetization data at ambient pressure.} Main panel: magnetization data collected on a powder sample as a function of temperature, both for 0.1 T and 3 T applied field. Data at low field were multiplied by a factor of 10. Inset: Field-dependent magnetization data collected at 2 K and 30 K.}\label{fig:S4}
    \end{figure}

    Temperature- and field-dependent magnetization data were collected on a powder sample at ambient pressure using a Quantum Design SQUID magnetometer. Low-field (0.1 T) temperature dependent data (main panel of Fig. \ref{fig:S4}) show antiferromagnetic (AFM) ordering with Néel temperature of 8 K, in good agreement with previous reports \cite{Wong2016,Yan2019PRM}. The 3 T data in the main panel is collected in the spin-polarized phase \cite{Vavilova2023PRB,Hu2024PRB} where magnetization is near saturation at low temperature. Magnetization measurements as a function of field are shown in the inset of Fig. \ref{fig:S4}. Comparison of ambient pressure magnetization to ambient pressure XMCD data allows scaling the pressure-dependent XMCD signals to absolute magnetization values.
    
    XAS/XMCD experiments used monochromatic radiation from a Si(111) double crystal monochromator. Radiation from higher undulator harmonics was rejected using a Si mirror at 3 mrad incidence angle combined with angular detuning of the second crystal in the monochromator. Linearly polarized radiation from the planar undulator was converted to circular polarization using a 180 $\mu$m-thick diamond C(111) phase plate operated in Bragg transmission geometry. XMCD data were collected in helicity switching mode whereby x-ray helicity is alternated at each point in energy scans using a piezoelectric actuator to deflect the phase plate away from the energy-dependent (111) Bragg condition by the angular offset required to generate radiation with high degree of circular polarization ($P_c >0.93$). Most XMCD measurements were done for two magnetic field directions, along and opposite the photon wavevector, which allows checking (and removing) any artifacts of non-magnetic origin in XMCD signals. XANES data is obtained by averaging over x-ray helicity.
    
    The integration of dipolar and quadrupolar XMCD signals (shown in Fig. 2e of main paper) was carried out over the [7.715-7.735] and [7.705-7.712] keV energy ranges, respectively. This separation allows comparing the response of Co $4p$ and Co $3d$ orbital contributions.
    
\begin{figure}[ht]
    \centering
        \includegraphics[width=\textwidth]{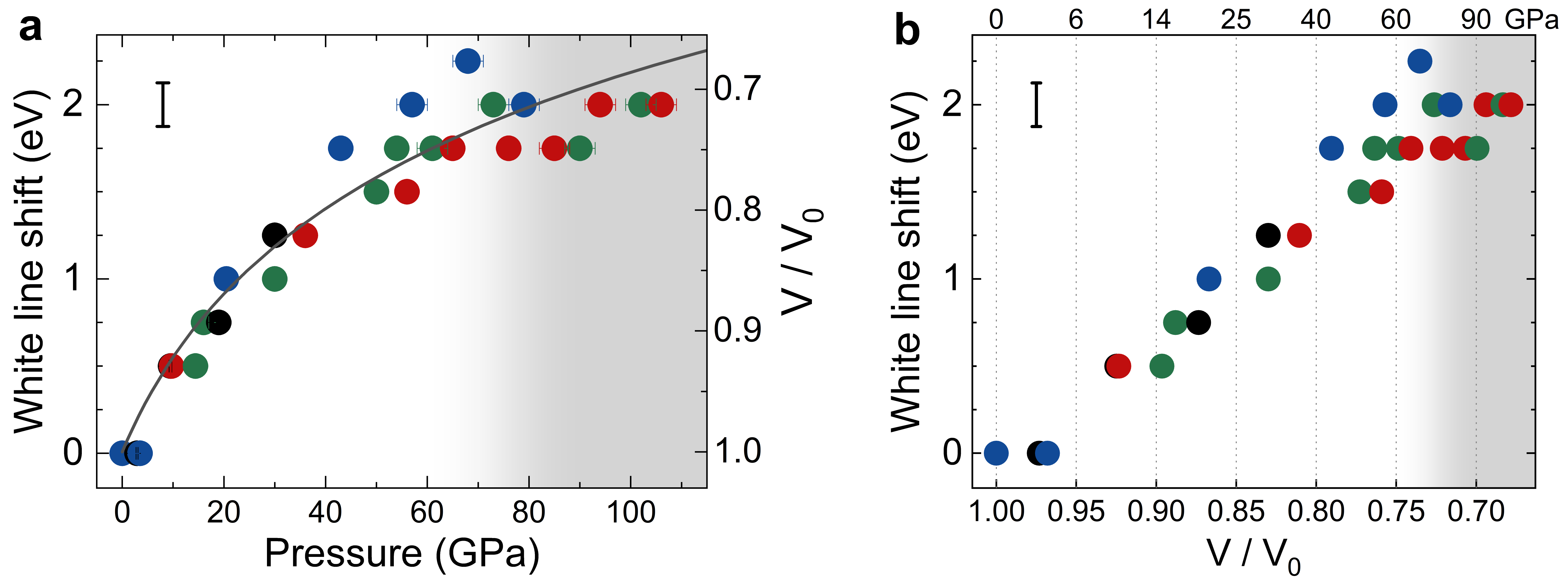}
        \caption{{\bf White line energy shift as a function of pressure and reduced volume.} \textbf{a} White line energy shift versus pressure alongside its dependence on reduced volume (right axis) obtained from the EOS derived from XRD (main paper Fig. 1b). \textbf{b} White line energy shift with respect to reduced volume. The top axis shows pressure values (rounded to integer values) obtained from the EOS. A typical error bar, dictated by the energy resolution of the monochromator, is shown in the upper left corner. White and gray regions correspond to Co HS and LS states, respectively.}\label{fig:S5}
    \end{figure}

    Figure \ref{fig:S5} shows the XAS white line energy shift as a function of both pressure and reduced volume. The Birch-Murnaghan equation of state (EOS) derived from XRD (shown in Fig. 1b) was used to derive the reduced volume. As discussed in the main paper, the energy shift tracks the volume contraction up to about 60-70 GPa. At higher pressures (smaller volume) the energy shift appears to slow down significantly in the region where the spin transition is observed.
    
\begin{figure}[ht]
    \centering
        \includegraphics[width=6.5 in]{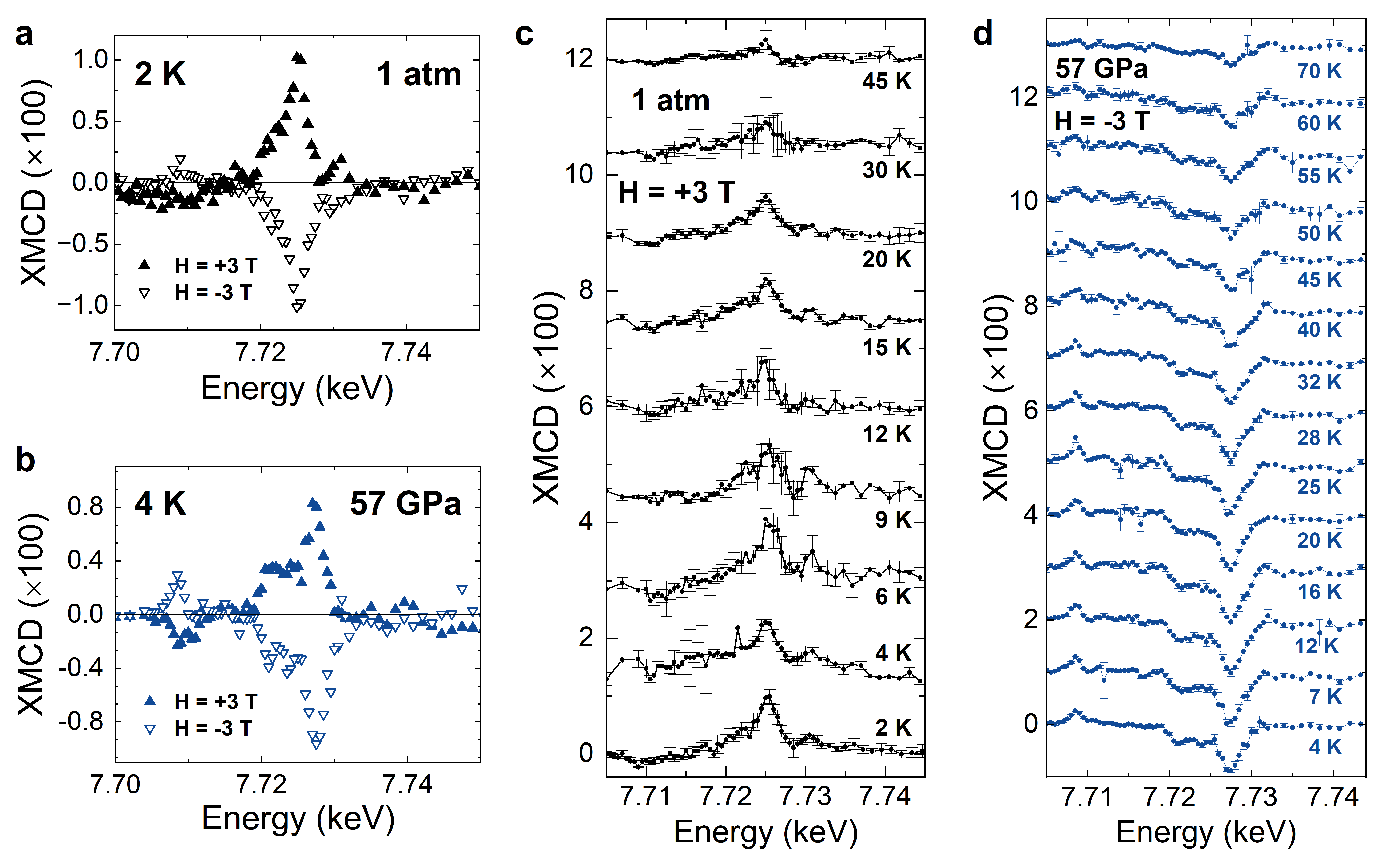}
        \caption{{\bf Temperature dependence of raw XMCD data.} XMCD spectra collected with magnetic field applied parallel to the photon wavevector (closed symbols, $H = +3$ T) and antiparallel to it (open symbols, $H = -3$ T) at \textbf{a} 1 atm, 2 K and \textbf{b} 57 GPa, 4 K. \textbf{c} Temperature dependence of ambient pressure XMCD obtained for $+3$ T and \textbf{d} Temperature dependence of XMCD spectra at 57 GPa for a $-3$ T field.}\label{fig:S6}
    \end{figure}

    Figure \ref{fig:S6} shows temperature-dependent XMCD data collected at 1 atm and 57 GPa. The good reversal of XMCD signal with applied field direction, both for quadrupolar and dipolar features, is also highlighted. Artifacts of non-magnetic origin are one order of magnitude smaller than the XMCD signal intensity. The data in Figs. \ref{fig:S6}c,d highlight the persistence of robust XMCD signal to much higher temperatures at 57 GPa compared to 1 atm. 
    
\begin{figure}[ht]
    \centering
        \includegraphics[width=6.5 in]{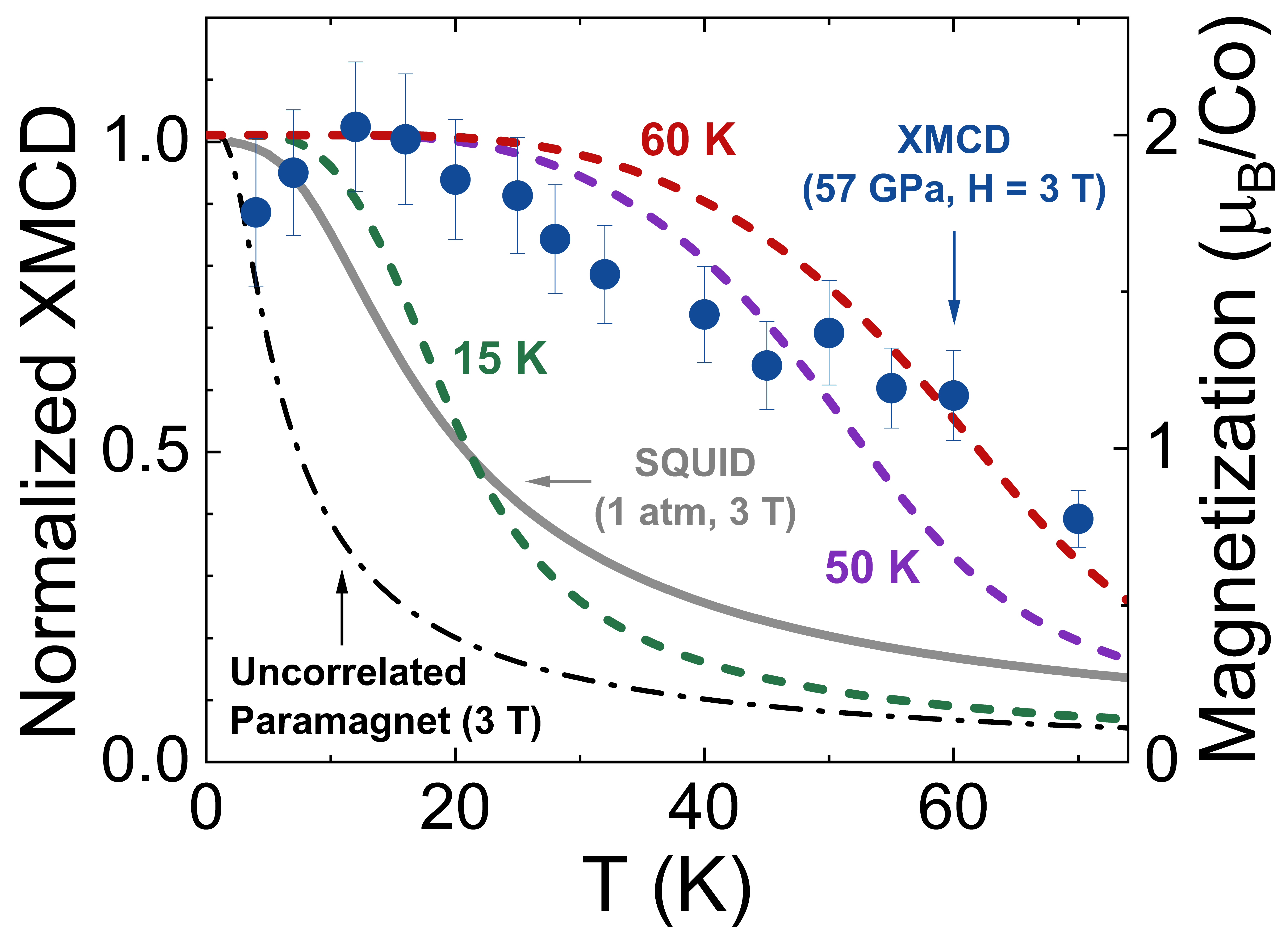}
        \caption{{\bf Modeling of temperature-dependent XMCD data under pressure.} Blue points and solid gray line show experimental data collected at 57 GPa (XMCD) and 1 atm (SQUID) respectively. These data were taken at 3 T and are the same as shown in Fig. 2f on the main paper. The calculated magnetization of an uncorrelated $S=1/2$ paramagnet at 3 T is shown with the black dash-and-dotted line. Models for the temperature dependence of magnetization for first nearest-neighbor ferromagnetic interactions of Ising spins are also shown for three different ordering temperatures, 15, 50 and 60 K, in green, purple and red dashed lines, respectively.}\label{fig:S7}
    \end{figure}

    Figure \ref{fig:S7} shows data displayed in Fig. 2f of the main paper together with the simulation of the magnetic response of an uncorrelated paramagnet, as well as of ferromagnetically-interacting (FM) Ising spins \cite{Brush1967}, both on the presence of an external magnetic field ($H=3$ T). These simulations considered the local moment to be $J=1/2$ and a powder-averaged gyromagnetic ratio of $g=4$, based on extrapolations from previous theoretical and experimental work \cite{Veenendaal2023,Liu2020PRL}. For a FM nearest-neighbor exchange coupling $J_1$, the Ising Hamiltonian of the system is given by
    
    \begin{equation}\label{eq:hamiltonian}
        \mathcal{H} = - J_1 \sum_{1NN} s_i s_j - g \mu_B J H \sum_{i=1}^N s_i,
    \end{equation}
    
    where $s_i$ is the spin moment of the $i$-th spin and $\mu_B$ is the Bohr magneton. Thus, the mean field magnetization in units of $\mu_B$ per Ising spin (in our case, one pseudospin per Co ion) is given by $M = g J \bar{s}$, with $\bar{s}$ being the mean spin moment. $\bar{s}$ can be calculated for a given field $H$ and temperature $T$ by iterating the equation below:
    
    \begin{equation}\label{eq:ising-spin}
        \bar{s}_{i+1} = \tanh \bigg[ \frac{T_c}{T} \bigg( \frac{H}{H_c} +\bar{s}_i \bigg) \bigg],
    \end{equation}
    
    where $T_c=\frac{nJ_1}{2k_B}$ is the ferromagnetic ordering temperature, $H_c = \frac{k_B T_c}{g \mu_B J}$ is the critical field, $n$ is the number of nearest-neighbor atoms, and $k_B$ is the Boltzmann constant. Similarly, the Hamiltonian for an uncorrelated paramagnet under an external field is given by the second term of equation \ref{eq:hamiltonian}. Therefore, the magnetization as a function of temperature can be calculated (in $\mu_B$ per spin) by:
    
    \begin{equation}\label{eq:uncorr-paramag}
        M = g J \tanh \bigg(\frac{g \mu_B J H}{k_B T} \bigg).
    \end{equation}
    
\begin{figure}[ht]
    \centering
        \includegraphics[width=5 in]{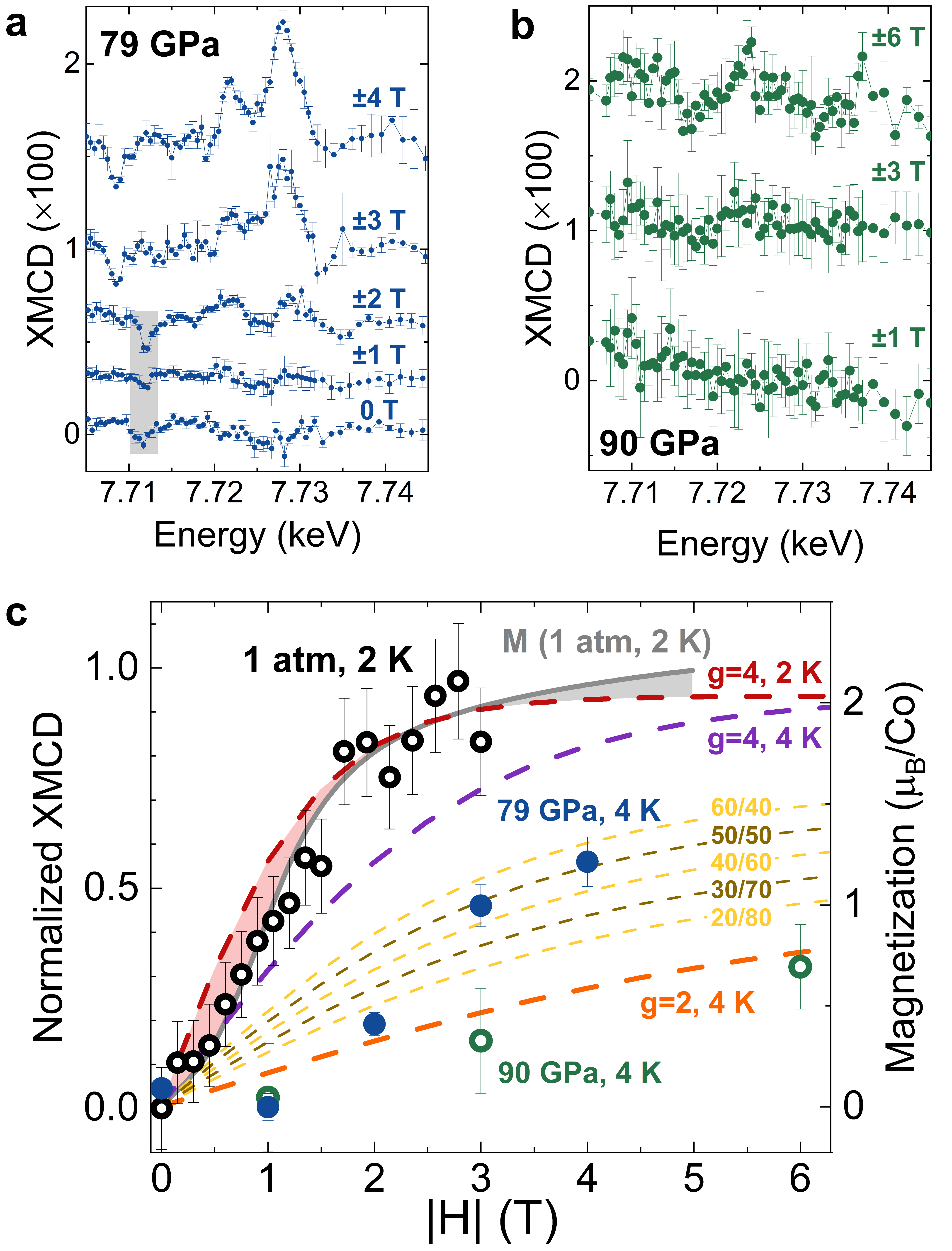}
        \caption{{\bf Magnetic field dependence of raw XMCD data.} Raw XMCD spectra as a function of magnetic field at \textbf{a} 79 GPa and \textbf{b} 90 GPa. The gray box in the left panel highlights a glitch from Bragg diffraction in diamond anvils which was not well compensated in those measurements when subtracting absorption spectra for opposite helicities. \textbf{c} XMCD (points), SQUID magnetization data (continuous gray line) and models of magnetization for an uncorrelated (Curie) paramagnet (dashed lines) as a function of magnetic field. Integrated XMCD, and magnetization data, are those presented in Fig. 2g of the main paper. Temperature and $g$ values are indicated for each model ($g\approx4$ is the powder average value in the HS, $J_{\rm eff}=\frac{1}{2}$ state \cite{Liu2020PRL,Veenendaal2023}, while $g=2$ is used for the LS, $J=S=\frac{1}{2}$ state). The shaded regions highlight the differences between the paramagnetic model (with $g=4$, $T=2$ K) and SQUID magnetization data at 1 atm; see text for details. Models corresponding to weighted averages of the HS $g=4$ and LS $g=2$ curves (both at $T=4$ K) with HS to LS fractions ranging from 60/40 to 20/80 are also shown.}\label{fig:S8}     
    \end{figure}

    The much reduced magnetization calculated for a classical paramagnet is clearly distinct from both the ambient and the 57 GPa data, confirming the presence of exchange correlations in the polarized phase at 3 T. While the Ising model does not properly describe the data, as expected for a field-induced polarized phase, it allows us to provide a crude estimate of a three-fold increase in ferromagnetic correlations between 1 atm and 57 GPa, 
    
    Figure \ref{fig:S8}a,b shows raw field-dependent data ($T=4$ K) used to obtain the integrated XMCD signals in Fig. 2g of the main paper. At 79 GPa, the pre-edge peak can only be seen in the 3 T and 4 T spectra once its intensity is large enough to be above noise level. At this pressure the XMCD signal in 2 T field is suppressed by a factor of 4 relative to 1 atm. At 90 GPa, the XMCD signal at 3 T has nearly vanished although a measurable signal is recovered in 6 T field. Field-dependent XMCD signal at 1 atm (shown in Fig. 2g of main paper) was collected at a fixed energy of 7.725 keV, at which the XMCD intensity is maximized, and therefore no raw energy-dependent XMCD data versus applied field is shown for 1 atm. Figure \ref{fig:S8}c shows data from Fig. 2g of the main paper together with the simulation of field-dependent magnetization curves (in $\mu_B$ per Co atom) for a classical paramagnet at a fixed temperature using the following equation,
    
    \begin{equation}\label{eq:brillouin}
        M = g \bigg[ \bigg(J+\frac{1}{2}\bigg) \coth\bigg(J+\frac{1}{2}\bigg)x - \frac{1}{2}\coth\frac{x}{2} \bigg]
    \end{equation}
    
    which is derived from the Brillouin function \cite{Darby1967}, with the dimensionless parameter $x = \frac{g \mu_B H}{k_B T}$.
    
    The paramagnetic model with $g=4$, $T=2$ K overestimates the 1 atm magnetization for $H < 1.5-2.0$ T due to the presence of magnetic order with antiferromagnetic correlations below this critical field (red shaded area) \cite{Gu2024PRB,Vavilova2023PRB,Hu2024PRB,Li2022PRX}. The model matches the data in the 2-3 T field region where a magnetically disordered, spin-liquid-like phase was reported\cite{Vavilova2023PRB,Hu2024PRB,Li2022PRX}. The model underestimates the data above this field, within the field-polarized phase\cite{Vavilova2023PRB,Hu2024PRB,Li2022PRX}, likely due to a small Van Vleck paramagnetic contribution. This contribution can be estimated by adding a linear, field-dependent term ($+\chi_{VV}H$) to equation \ref{eq:brillouin}, such that the Van Vleck magnetic susceptibility is estimated to be $\chi_{VV}=0.030(3)$ $\mu_B$/T per Co$^{2+}$ ion, which is comparable to values found for other high spin Co$^{2+}$ systems in octahedral coordination \cite{Haraguchi2024arxiv,Shirata2012PRL}.
    
    XMCD experiments at pressures above 9.6 GPa were carried out at $T=4$ K to avoid freezing of pressurized Helium in compression and decompression membranes. The high-spin (HS) to low-spin (LS) transition at about 70 GPa converts the spin-orbit entangled pseudospin moment $J_{\rm eff}=|L-S|=|1-3/2|=1/2$, with powder-averaged gyromagnetic ratio $g=4$, into a spin-only moment ($L=0$, $J=S=1/2$ and $g=2$).  
    
    The field-dependent XMCD data at $P=79$ GPa cannot be modeled by any of the curves in Fig. \ref{fig:S8}c, possibly due to proximity to the spin transition and presence of both HS and LS components, as seen by a saturation magnetization at high fields intermediate between 1 atm (HS) and 90 GPa (LS) data. A 50/50 weighted average between models with gyromagnetic ratio $g=4$ and $g=2$ appears to match the high-field region of the 79 GPa data, as may be expected for a moment in transition from HS ($g=4$) to LS ($g=2$) state. Both 79 GPa and 90 GPa XMCD data have sizable error bars at low-field ($H=1$ T) which are comparable to the integrated XMCD signal itself, making it challenging to assert whether a gap for magnetic excitations is present. The paramagnetic model provides a good description of the 90 GPa data. Since IR spectroscopy shows that this cobaltate remains an insulator to at least 110 GPa, exchange interactions between localized LS, $S=\frac{1}{2}$ moments appear to be frustrated, potentially rendering the LS high-pressure state of this 2D honeycomb lattice a classical or quantum spin liquid.
    
\section*{X-ray emission spectroscopy}   

\begin{figure}[ht]
    \centering
        \includegraphics[width=6.5 in]{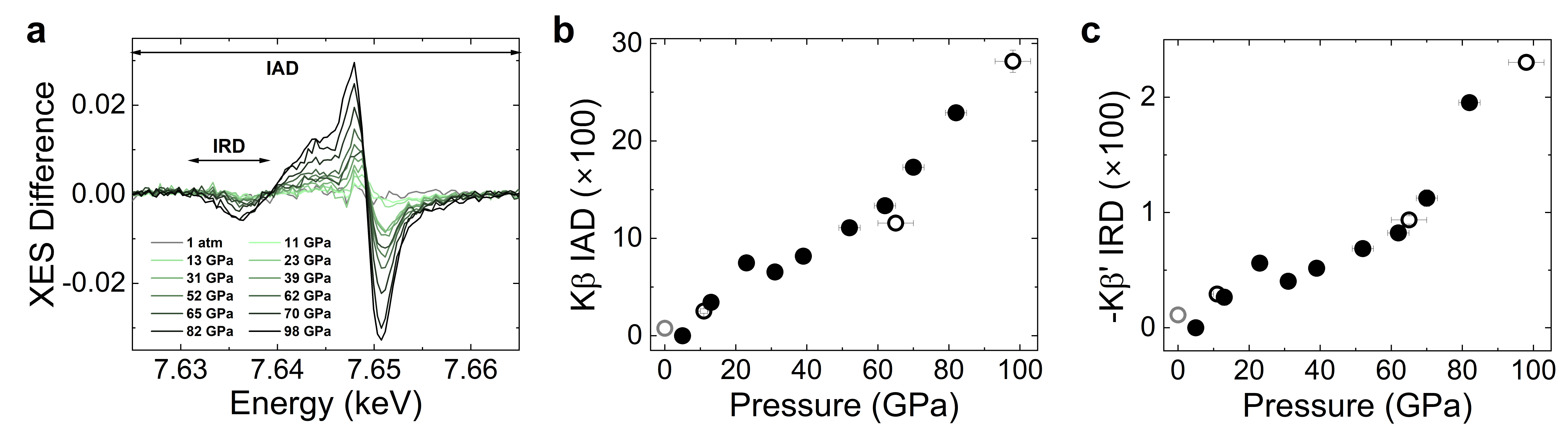} \caption{{\bf Integrals of difference XES spectra.} \textbf{a} Difference in emission spectra using 5 GPa spectra as reference. Horizontal arrows indicate energy regions for IAD and IRD integrations. \textbf{b} Pressure dependence of Co K$\beta$ IAD (7.609-7.675 keV) where difference spectra are integrated over the full energy range of x-ray emission; and \textbf{c} -K$\beta^{\prime}$ IRD, which integrates over the 7.631-7.639 keV energy range only. Closed circles are 300 K data and open circles are low temperature (30 K, 200 K) data. The 65 GPa point was collected during decompression. 1 atm, 300 K data measured outside of the DAC are also shown.}\label{fig:S9}
    \end{figure}

\begin{figure} [ht]
    \centering
        \includegraphics[width=\textwidth]{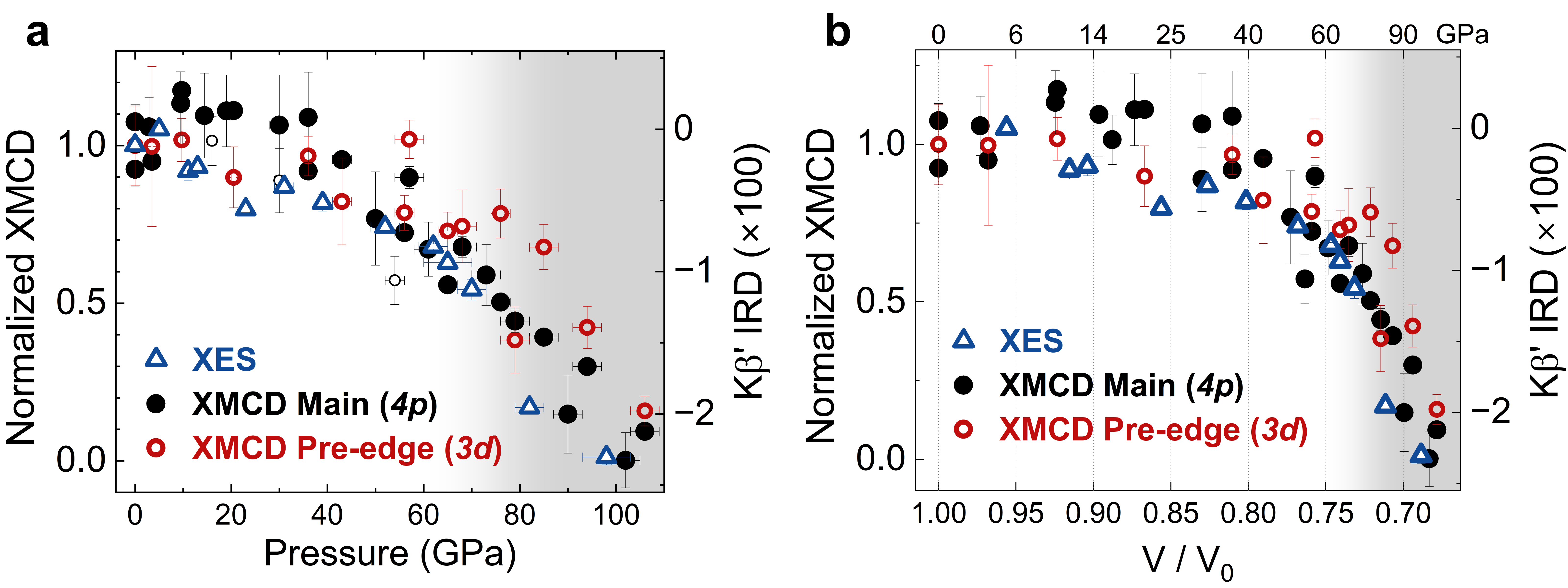}
        \caption{\textbf{Evolution of XMCD and XES intensity with pressure and reduced volume} Integrated XMCD and Co K$\beta'$ IRD as a function of \textbf{a} pressure and \textbf{b} reduced volume. Pressure values approximated to the unit are shown on the top axis in (b). The suppression of magnetic susceptibility above about 60 GPa is correlated with changes in local moment across a HS (white region) to LS (gray region) transition.}\label{fig:S10}
    \end{figure}

    XES data analysis was carried out using both the integrated absolute difference (IAD) methods \cite{Vanko2006} and the integrated relative difference (IRD) method \cite{Mao2014AmMin,Li2019PRB}. In both cases emission spectra are normalized to an area of unity and differences calculated by subtracting the 5 GPa spectrum as the reference spectrum. These difference spectra are shown in Fig. \ref{fig:S9}a. In the IAD method, difference spectra were integrated over the entire energy range (7.609-7.675 keV) with Fig. \ref{fig:S9}b showing the pressure-dependence of these integrated IAD values. In contrast, the IRD method uses integrals over the $K\beta'$ peak only, from 7631.4 to 7639.2 eV. Figure \ref{fig:S9}c shows integrated IRD values (same as in the main paper Fig. 3e, multiplied by $-1$ for better comparison with panel b). By comparing panels b and c, one notes that the HS-LS transition is clearly seen in the IRD data but manifests a more subtle slope change in the IAD data. This is because the IAD data is dominated by a large contribution from the $K\beta_{1,3}$ main line moving to lower energies as a result of diminishing $3p-3d$ intra-atomic exchange interaction upon compression driven by $3d$ delocalization. The $K\beta^{\prime}$ line is more sensitive to the magnitude of the local moment. We attribute the slow reduction in IRD intensity below the spin transition to the compression-driven increase in Co $3d -$O $2p$ covalency/hybridization. Figure \ref{fig:S10} compares the evolution of XMCD and XES intensity with pressure as well as with reduced volume, the latter derived from the Birch-Murnaghan EOS obtained from XRD (Fig. 1b). While a local moment persists in the LS, insulating phase, XMCD shows a vanishing spin polarization in $H=3$ T applied field at 100 GPa ($T=4$ K).
    
\section*{Density Functional Theory Calculations}

\begin{figure}[ht]
    \centering
        \includegraphics[width=6.5 in]{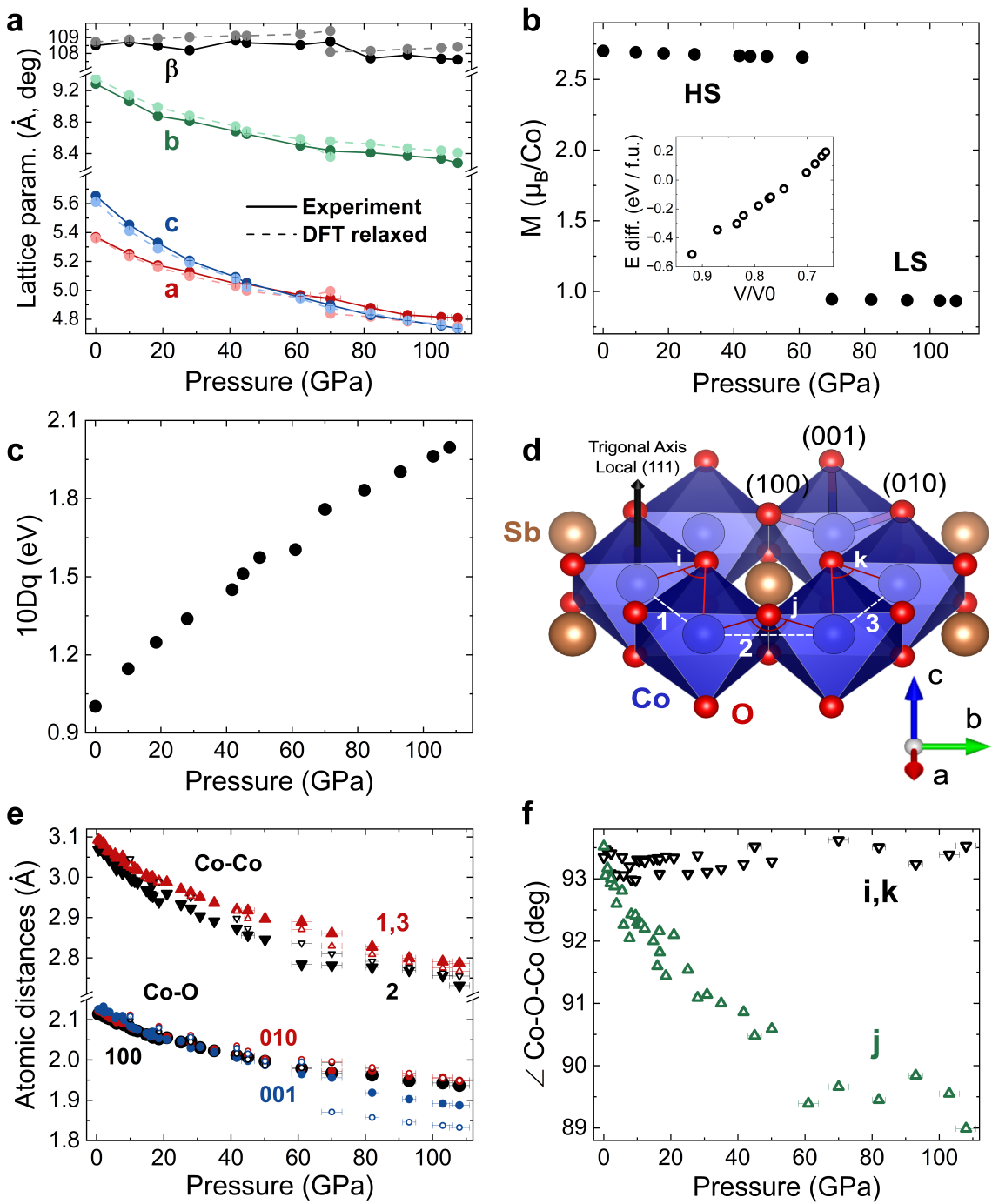}
        \caption{{\bf Spin transition, interatomic distances, and bonding angles derived from DFT calculations of relaxed structures.} \textbf{a} Comparison between experimental and DFT-relaxed lattice parameters. \textbf{b} Magnetic moment and \textbf{c} $10Dq$ crystal field calculated after DFT-relaxation of atomic positions and lattice parameters. The inset on panel \textbf{b} shows the energy difference between HS and LS states as a function of reduced volume. \textbf{d} Sb-Co-O honeycomb plane including labels for Co-Co and Co-O distances, as well as Co-O-Co angles, used in panels (e,f) of this figure. Crystalline $a$, $b$ and $c$-axis and trigonal distortion vector are also shown for reference. \textbf{e} Octahedral Co-O distances (circles) and Co-Co distances in the honeycomb plane (up and down triangles). Out-of-plane Co-Co distance matches values for the $c$ lattice parameter. Closed symbols correspond to DFT theoretical structures using experimental lattice parameters but relaxed atomic positions, while open circles correspond to all-relaxed structures (lattice parameters relaxed). \textbf{f} Co-O-Co angles in edge-shared octahedra. When not shown, error bars are smaller than symbols.}\label{fig:S11}
    \end{figure}

    Density Functional Theory plus U (DFT+U) calculations of relaxed structures were performed in two independent ways: (1) constraining lattice parameters to experimental values and relaxing atomic positions, and (2) relaxing both lattice parameters and atomic positions while keeping the unit cell volume constrained to experimental values. For the latter calculations, the starting point lattice parameters were those derived from XRD. Reasonable agreement is found between relaxed DFT lattice parameters and experimental values as can be seen in Fig. \ref{fig:S11}a. Figure \ref{fig:S11}b shows the calculated magnetic moment, $M$, across the HS to LS transition for fully relaxed DFT structures. The pressure where this transition takes place is in good agreement with results from x-ray emission spectroscopy experiments and cluster calculations. The inset on Fig. \ref{fig:S11}b shows the calculated energy difference between HS and LS states. DFT calculations also show that the $t_{2g}-e_g$ crystal field splitting, 10$Dq$, increases almost monotonically over the entire pressure range, with a slight change in slope at the spin transition (Fig. \ref{fig:S11}c). The energy levels of $t_{2g}$ and $e_g$ orbitals are obtained by projection of DFT bands to Co $d$ Wannier orbitals. The increased crystal field upon compression is the driving force of the HS-LS transition. DFT (cluster) calculations show the spin transition taking place at $10Dq=1.8$ eV (2.27 eV), respectively, comparable to threshold values reported for MnO \cite{Mattila2007}. Co ions retain the localized nature of their moments up to 1 Mbar, although covalency increases upon compression as evidenced by the non-negligible change in $K\beta^{\prime}$ IRD below 70 GPa (see Fig. \ref{fig:S9}).
    
    Figure \ref{fig:S11}e shows Co-O and Co-Co distances obtained from DFT calculations of relaxed structures. Below 70 GPa, Co-O distances in all 3 local directions decrease at the same rate but, at the spin transition, a splitting of distances is observed between (001) and (100)/(010) local axes directions (see Fig. \ref{fig:S11}d for local coordinate axes notation). The size of the splitting differs between calculations where only atomic positions are relaxed, versus fully relaxed structures. This splitting is consistent with a Jahn-Teller distortion being present in CoO$_6$ octahedra surrounding the LS Co$^{2+}$ ions. The choice of local coordinates follows the notation used in previous work \cite{Veenendaal2023}. For Co-Co distances, a similar P-dependent behavior is observed for all 3 bond directions, independent of relaxation method, with more data scatter seen for structures that constrained lattice parameters to experimental values. Figure \ref{fig:S11}f shows Co-O-Co angles across edge shared octahedra based on calculations using experimental lattice parameters and relaxed atomic positions. The angle involving two Co ions along the $b$-axis, j, is reduced by $\sim$ 5\% at 108 GPa from its ambient pressure value, while the other 2 degenerate angles remain close to their starting values of 93.5$^{\circ}$. Only the j angle approaches the ideal 90$^{\circ}$ bonding expected for edge-shared regular octahedra. Figure \ref{fig:S11}d summarizes notation used to label all atomic distances and angles shown in panels e,f.

\section*{Infrared Absorption Spectroscopy}

\begin{figure} [ht]
    \centering
        \includegraphics[width=6.5 in]{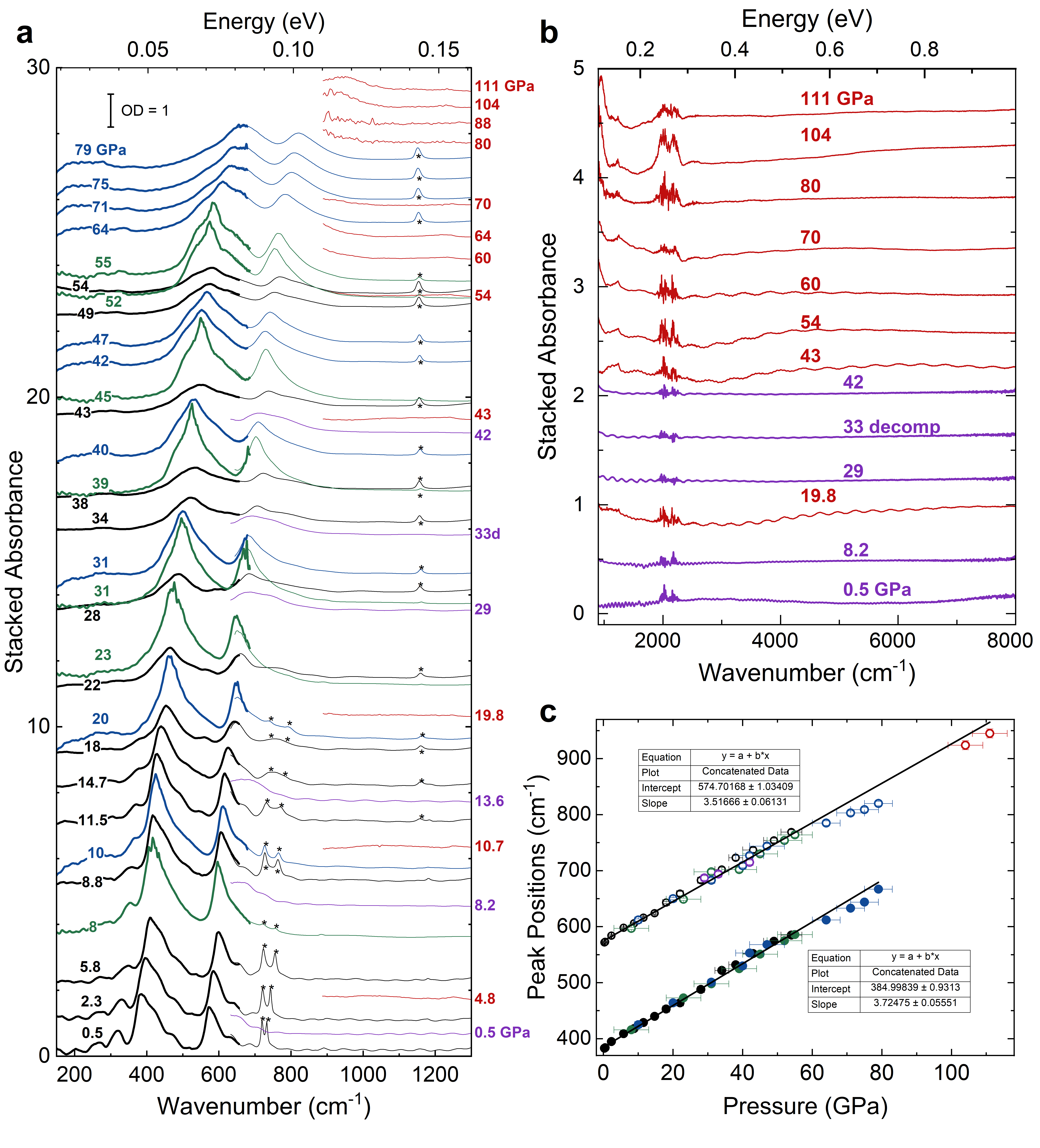}
        \caption{{\bf FIR and MIR absorption spectra, phonon hardening.}  \textbf{a} Infrared absorbance spectra (stacked) as a function of pressure (in GPa). \textbf{b} Extended MIR data from runs in which KBr was used as pressure transmitting medium. \textbf{c} Pressure dependence of the two most intense phonon peaks shown on panel a. Different colors, used consistently in all panels, denote different experimental runs.}\label{fig:S12}
    \end{figure}

    Single crystal infrared absorption spectra both in the far infrared (FIR: $<650$ cm$^{-1}$) and mid infrared (MIR: $650-8000$ cm$^{-1}$) regions are shown in Fig. \ref{fig:S12}. High-pressure absorbance was measured in multiple runs using various DAC loadings. Three experiments focusing in the FIR region used petrol jelly as pressure medium, while two other experiments focused on MIR region used KBr. Petrol jelly has absorption peaks in the MIR and no energy-dependent absorption in the FIR, while KBr is transparent in the MIR but absorbs in the FIR \cite{Stolen1965PhysRev,Johnson1969PhysRev}. Absorbance is defined as $A=$ log$(I_0/I)$, where $I_0$ is the reference spectrum or background intensity, and $I$ is the intensity of infrared light transmitted through the sample. Note that FIR and MIR absorption measurements require different experimental setups (optics, detectors). Due to these constraints, $I_0$ is collected differently for these two regions: in the FIR, the background is measured as the transmission through diamond anvils (measured once at 1 atm before the sample is loaded), while the MIR reference spectrum is measured through both diamonds and KBr at each pressure point. 

    The lowest energy region ($< 40-60$ meV) in Fig. \ref{fig:S12} shows a flat energy dependence as a function of pressure up to 79 GPa. Metals show perfect reflectivity below their plasmon energy, usually $\sim$10 eV, so absorption experiments performed in transmission geometry would show vanishing transmission and diverging absorbance. Therefore, in the case of a pressure-dependent insulator-to-metal transition one would expect a Drude feature to develop at the lowest wavenumbers. The absence of such a feature in Fig. \ref{fig:S12} rules out metallization up to 79 GPa. Moreover, the only excitations seen in the FIR region are due to phonons which, for a metal, would end up completely buried by the Drude feature up to the plasmon edge. To extend our conclusion that the insulating state persists to even higher pressures, one can use the flatness of the MIR region. Metallization is accompanied by a reduction and eventual closure of an electronic gap (here a charge-transfer gap, $\sim 2-3$ eV at ambient pressure). An absorption peak associated with excitation across the charge-transfer gap would decrease in energy from the optical into the IR regime, as pressure increases. This is not observed until at least 111 GPa (Fig. \ref{fig:S12}b) allowing us to put a lower limit of $\sim 1$ eV for the charge transfer gap at this pressure. The lowest energy feature seen at high pressures in b is one of the main phonon peaks moving up from the FIR into the MIR region. 

    The background compensation for some pressure points in Fig. \ref{fig:S12}b was not perfect during the runs with KBr. This is most likely due to the various small apertures the beam goes through ending up being clipped by the gasket and not providing a true $I_0$ measurement for that particular run. This can be seen for example in the 104 GPa spectrum, where a diamond peak around 2000 cm$^{-1}$ appears. Figure \ref{fig:S12}c quantifies the hardening of the two primary phonons seen in the FIR spectra, one of which is also captured in the MIR spectra at the highest pressures of 104 GPa and 111 GPa. The linear dependence of the phonon hardening, together with absence of new phonon modes detected in the IR spectra, provide further support for the lack of structural transitions into the 1 Mbar range, consistent with the conclusions derived from XRD data.


%